\documentclass{aa}
\usepackage{graphicx}
\usepackage{txfonts}
\begin{document}
\title{
High angular resolution $N$-band observation of the silicate carbon star 
IRAS08002-3803 with the VLTI/MIDI instrument
\thanks{Based on observations made with the Very Large Telescope 
Interferometer of the European Southern Observatory. 
Program ID: 073.A-9002(A)}
}
\subtitle{Dusty environment spatially resolved}

\author{K.~Ohnaka\inst{1}, 
T.~Driebe\inst{1}, 
K.-H.~Hofmann\inst{1}, 
Ch.~Leinert\inst{2},
S.~Morel\inst{3}, 
F.~Paresce\inst{4},
Th.~Preibisch\inst{1}, 
A.~Richichi\inst{4},
D.~Schertl\inst{1}, 
M.~Sch\"oller\inst{3}, 
L.~B.~F.~M.~Waters\inst{5}, 
G.~Weigelt\inst{1}, 
M.~Wittkowski\inst{4}
}

\offprints{K.~Ohnaka}

\institute{
Max-Planck-Institut f\"{u}r Radioastronomie, 
Auf dem H\"{u}gel 69, 53121 Bonn, Germany\\
\email{kohnaka@mpifr-bonn.mpg.de}
\and
Max-Planck-Institut f\"ur Astronomie, 
K\"onigstuhl 17, 69117 Heidelberg, Germany
\and
European Southern Observatory, Casilla 19001, Santiago 19, Chile
\and
European Southern Observatory, Karl-Schwarzschild-Str.~2, 
85748 Garching, Germany
\and
Astronomical Institute ``Anton Pannekoek'', Kruislaan 403, 
1098 SJ Amsterdam, The Netherlands
}

\date{Received / Accepted }

\abstract{
We present the results of $N$-band spectro-interferometric 
observations of the silicate carbon star IRAS08002-3803 with the 
MID-infrared Interferometric instrument (MIDI) at the 
Very Large Telescope Interferometer (VLTI) of the European 
Southern Observatory (ESO).  
The observations were carried out using two unit telescopes 
(UT2 and UT3) with projected baseline lengths ranging from 39 to 47~m.  
Our observations of IRAS08002-3803 
have spatially resolved the dusty environment of a 
silicate carbon star for the first time and revealed an 
unexpected wavelength dependence of the angular size in the $N$ band: 
the uniform-disk diameter is found to be constant and $\sim$36~mas 
(72~\mbox{$R_{\star}$}) between 8 and 10~\mbox{$\mu$m}, while it steeply increases 
longward of 10~\mbox{$\mu$m}\ to reach $\sim$53~mas (106~\mbox{$R_{\star}$}) 
at 13~\mbox{$\mu$m}.  
Model calculations with our Monte Carlo radiative transfer code show 
that neither spherical shell models nor axisymmetric disk 
models consisting of silicate grains alone can simultaneously 
explain the observed wavelength dependence of the visibility and 
the spectral energy distribution (SED).  
We propose that the circumstellar environment of IRAS08002-3803 
may consist of two grain species coexisting in the disk: 
silicate and a second grain species, 
for which we consider amorphous carbon, large silicate grains, and 
metallic iron grains.  
Comparison of the observed visibilities and SED with our models 
shows that such disk models can fairly --- though not entirely 
satisfactorily --- reproduce the observed SED and 
$N$-band visibilities.  
Our MIDI observations and the 
radiative transfer calculations lend support to the picture where 
oxygen-rich material around IRAS08002-3803 is stored in a circumbinary 
disk surrounding the carbon-rich primary star and its 
putative low-luminosity companion.  
}   %  END OF ABSTRACT

\titlerunning{VLTI/MIDI observation of the silicate carbon star IRAS08002-3803}
\authorrunning{Ohnaka et al.}
\maketitle

\section{Introduction}
\label{sect_intro}

Slow but massive mass loss at the asymptotic giant branch (AGB) leads 
to the formation of circumstellar dust envelopes.  The dust species 
formed in the circumstellar envelope reflect the chemical composition 
of the photosphere, that is, dust grains such as silicate 
and corundum (Al$_2$O$_3$) are observed around oxygen-rich stars 
(M giants),  while amorphous carbon and/or silicon carbide (SiC) are 
observed around carbon stars.  

Therefore, the discovery of carbon stars showing (amorphous) silicate 
emission in the IRAS Low Resolution Spectra (LRS) by Little-Marenin 
(\cite{little-marenin86}) and Willems \& de Jong (\cite{willems86}) 
struck stellar spectroscopists as baffling, and the origin of these 
``silicate carbon stars'' is still a puzzle to date.  
The spectroscopic studies which followed this discovery revealed that 
the optical and near-infrared spectra of silicate carbon stars are 
dominated by strong 
absorption due to C$_2$ and CN, confirming that the photospheres of 
the silicate carbon stars are indeed carbon-rich and they can be 
classified as J-type carbon stars, which are enriched in 
$^{13}$C (e.g., Lloyd-Evans \cite{lloyd-evans90}; 
Skinner et al. \cite{skinner90}; Chan \cite{chan93}). 
In fact, quantitative spectral analyses show that these silicate 
carbon stars have 
$^{12}$C/$^{13}$C ratios as low as 4--5 (e.g., Ohnaka \& Tsuji 
\cite{ohnaka99}; Abia \& Isern \cite{abia00}), 
which poses a great challenge for stellar evolution 
theory.  On the other hand, detection of H$_2$O and OH maser emission 
toward silicate carbon stars 
confirmed that they are associated with oxygen-rich 
circumstellar material (e.g., Nakada et al. \cite{nakada87}, 
\cite{nakada88}; Benson \& Little-Marenin \cite{benson87}; 
Little-Marenin et al. \cite{little-marenin88}; 
Barnbaum et al. \cite{barnbaum91}; Engels \cite{engels94a}).  

Willems \& de Jong (\cite{willems86}) and Chan \& Kwok 
(\cite{chan91}) suggested that silicate carbon stars are objects in 
transition from M giants to carbon stars.  However, this scenario is deemed 
to be unlikely as Lloyd-Evans (\cite{lloyd-evans90}) argues: 
the time scale for such a transitional 
object to be observed as a silicate carbon star is 
predicted to be a few decades, while some of the silicate carbon stars 
are known to have carbon-rich photospheres for 50 years 
(Little-Marenin et al. \cite{little-marenin87}).  
Furthermore, the infrared spectrum of the silicate carbon star 
V778~Cyg observed by Yamamura et al. (\cite{yamamura00}) with 
the Infrared Space Observatory (ISO) reveals that the silicate 
features at 10 and 18~\mbox{$\mu$m}\ exhibit no temporal variation 
14 years after IRAS LRS, which makes it unlikely that silicate carbon 
stars are short-lived transitional objects.  
Little-Marenin (\cite{little-marenin86}) proposed that silicate 
carbon stars are binaries consisting of a carbon star and an M giant, 
but this scenario has now also been rejected, because near-infrared 
spectroscopy (Lambert et al. \cite{lambert90}) as well as near-infrared 
speckle interferometry (Engels \& Leinert \cite{engels94}) 
detected no hint of an M giant companion toward silicate carbon stars.  

At the moment, the most widely accepted hypothesis suggests that 
silicate carbon stars have a low-luminosity companion (instead of 
an M giant companion), 
and that oxygen-rich material was shed by mass loss when the 
primary star was an M giant and this oxygen-rich material 
is stored in a circumbinary disk (Morris \cite{morris87}; 
Lloyd-Evans \cite{lloyd-evans90}) or in a circumstellar disk 
around the companion (Yamamura et al. \cite{yamamura00}) 
until the primary star becomes a carbon star.  
In fact, there is observational evidence of binarity and 
the presence of such a disk for some--if not all--silicate 
carbon stars.  
Lloyd-Evans (\cite{lloyd-evans90}) argues against the 
shell geometry based on the extinction in the optical as well 
as in the near-infrared and the amount of the far-infrared excesses 
found in some silicate carbon stars (including IRAS08002-3803 
studied here).  
Radial velocity measurements of two silicate carbon stars 
by Barnbaum (\cite{barnbaum91}) are consistent with motion 
in a binary system.  
Evidence for the presence of a low-luminosity companion 
surrounded by an accretion disk was found in the violet spectrum 
of the silicate carbon star BM~Gem (Izumiura \cite{izumiura03}).  
The presence of long-lived reservoirs of orbiting gas is 
inferred from the narrow CO emission lines ($J$ = 1 -- 0 and 
$J$ = 2 -- 1) obtained by Kahane et al. (\cite{kahane98}) and 
Jura et al. (\cite{jura99}).  
Recent high-resolution 22~GHz H$_2$O maser maps toward 
V778~Cyg obtained by 
Szczerba et al. (\cite{szczerba05}) and Engels (priv. comm.) 
also suggest the existence of a rotating disk. 

However, the geometry of the reservoir of oxygen-rich material 
is still controversial.  Based on the ISO spectrum of V778~Cyg, 
Yamamura et al. (\cite{yamamura00}) argue that the oxygen-rich 
dust is stored in a circumstellar disk around the companion.  
On the other hand, Molster et al. (\cite{molster99},
\cite{molster01}) identified 
pronounced crystalline silicate emission in the 
silicate carbon star IRAS09425-6040 observed with ISO and 
postulate that crystalline silicate dust forms in a dense, 
circumbinary disk.  
Thus, it is not yet clear whether or not only one of the above 
two scenarios--circumbinary or circum-companion disk--applies 
to all silicate carbon stars.   

Observations with high spatial resolution within 
the silicate emission feature would be a most 
direct approach for investigating the dusty environment of silicate 
carbon stars.  
The MIDI instrument at the VLTI provides us with an excellent opportunity 
to directly study the circumstellar environment of silicate carbon stars 
in the 10~\mbox{$\mu$m}\ region, exactly where silicate emission from 
the oxygen-rich reservoir is located.  

In this paper, we present the results of $N$-band spectro-interferometric 
observations of the silicate carbon star IRAS08002-3803 
(Hen~38, GCCCS~1003\footnote{A General Catalogue of Cool Carbon Stars,
  Stephenson (\cite{stephenson73})}, GCCGCS~2011\footnote{A General 
Catalogue of Cool Galactic Carbon Stars, Stephenson
(\cite{stephenson89})}, hereafter IRAS08002), 
using the VLTI/MIDI instrument.  
The IRAS LRS of IRAS08002 shows the prominent silicate emission features 
at 10~\mbox{$\mu$m}\ as well as at $\sim$18~\mbox{$\mu$m}\ with a 12~\mbox{$\mu$m}\ flux 
of 40.3~Jy.  While most of the well studied silicate carbon stars 
are in the northern sky, IRAS08002 
is the brightest one among the few silicate carbon stars observable 
from the VLTI.  Lloyd-Evans (\cite{lloyd-evans90}) notes 
that IRAS08002 shows a 10~\mbox{$\mu$m}\ 
feature much broader than that observed in silicate carbon stars 
such as V778~Cyg and BM~Gem.  
Kwok \& Chan (\cite{kwok93}) also mention that some silicate carbon
stars (including IRAS08002) show silicate emission features broader 
than usual oxygen-rich AGB stars, 
while others show silicate features as narrow as in usual AGB stars.  
No H$_2$O maser is detected toward IRAS08002 
(Nakada et al. \cite{nakada88}; Deguchi et al. \cite{deguchi90}), 
and no radial velocity monitoring observations to confirm its 
binary nature are available in the literature.  

In Sect.~\ref{sect_obs}, we present the observational results. 
In Sect.~\ref{sect_code}, we describe our 
radiative transfer code based on the Monte Carlo technique.
The results of our model calculations for IRAS08002 and the 
comparison with the observations are presented in 
Sect.~\ref{sect_modeling} and \ref{sect_multgrain}. 
We discuss possible scenarios for the circumstellar environment 
of IRAS08002 in Sect.~\ref{sect_discuss}.

\section{MIDI observations}
\label{sect_obs}

\begin{table}
\begin{center}
\caption {Summary of the MIDI observations of IRAS08002: 
date, modified Julian Date (MJD), time of observation (Universal Time=UTC), 
projected baseline length $B_{\rm p}$, and 
position angle of the projected baseline on the sky (P.A.).
}
\vspace*{-2mm}

\begin{tabular}{r c c c r r}\hline
\# & Date & MJD & $t_{\rm obs}$ (UTC) & $B_{\rm p}$ (m) & P.A. (\degr)\\ 
\hline
1        & 2004 Feb. 09 & 53045.135 & 03:15:02       &   46.05        &   39.48\\
2        & 2004 Feb. 09 & 53045.248 & 05:56:56       &   39.11        &   56.96\\
3        & 2004 Feb. 10 & 53046.137 & 03:17:44       &   45.94        &   40.35\\
4        & 2004 Feb. 11 & 53047.093 & 02:13:32       &   46.58        &   32.03\\
\hline
\label{table_obs}
\vspace*{-7mm}

\end{tabular}
\end{center}
\end{table}

\begin{table}
\begin{center}
\caption {
List of calibrators used in the present work, 
together with 12~\mbox{$\mu$m}\ fluxes ($F_{12}$), 
uniform-disk diameters ($d_{\rm{UD}}$) 
and the date as well as the time stamp ($t_{\rm obs}$) 
of the MIDI observations. 
The uniform-disk diameters were taken from the CalVin 
list available at ESO
(http://www.eso.org/observing/etc/). 
}
\begin{tabular}{l r r c l}\hline
Calibrator & $F_{12}$ (Jy) & $d_{\rm{UD}}$ (mas)  & Date & $t_{\rm obs}$ (UTC) \\ 
HD number  &               &                      & (2004)&                     \\ \hline \hline
18322      & 13.3          & $2.50\pm 0.12$       & Feb. 09 &  00:39:11 \\ 
           &               &                      & Feb. 09 &  01:39:04 \\ \hline
49161      & 10.4          & $2.88\pm 0.17$       & Feb. 09 &  02:45:52 \\ 
           &               &                      & Feb. 10 &  04:43:21 \\ \hline
67582      & 9.3           & $2.69\pm 0.25$       & Feb. 09 &  06:24:15 \\ 
           &               &                      & Feb. 09 &  07:13:39 \\ 
           &               &                      & Feb. 10 &  02:37:04 \\ 
           &               &                      & Feb. 10 &  03:46:12 \\ \hline
107446     & 32.4          & $4.54\pm 0.23$       & Feb. 09 &  08:08:53 \\ 
           &               &                      & Feb. 10 &  08:16:44 \\ \hline
85951      & 12.6          & $3.48\pm 0.18$       & Feb. 11 &  02:42:15 \\ 
           &               &                      & Feb. 11 &  05:37:52 \\ \hline
120404     & 13.3          & $3.03\pm 0.24$       & Feb. 11 &  07:36:34 \\ 
           &               &                      & Feb. 11 &  08:27:38 \\ \hline
\label{table_calib}
\end{tabular}
\end{center}
\end{table}

\begin{figure*}[!hbt]
\sidecaption
\includegraphics[width=12cm]{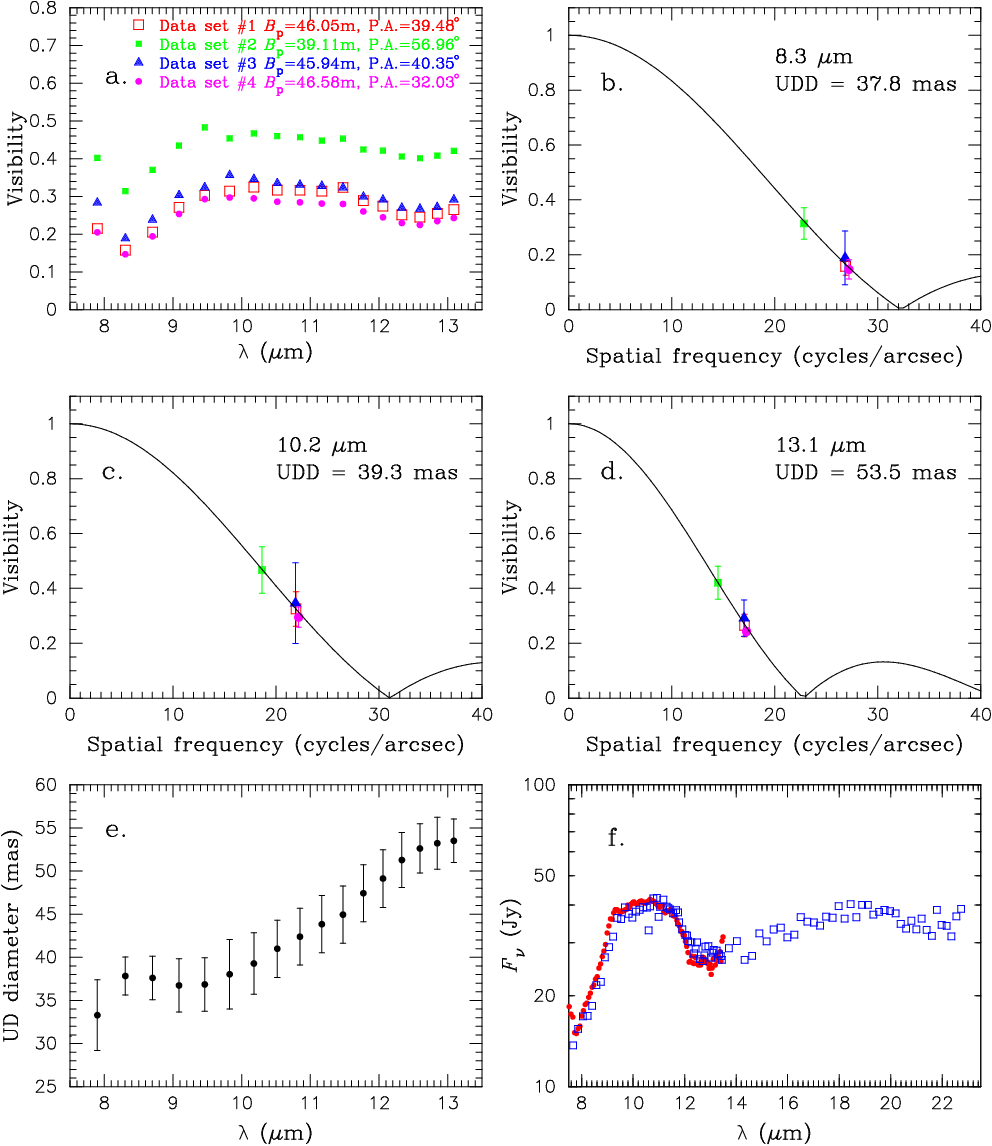}
\caption{$N$-band visibilities and spectra observed for IRAS08002. 
{\bf a:} Observed visibilities plotted as a function of wavelength.  
The errors of 
the observed visibilities are typically $\pm 10$--15\%, but the error 
bars are omitted in this panel for the sake of visual clarity.  
{\bf b--d:} Visibilities at three representative wavelengths in the 
$N$ band are plotted as a function of spatial frequency, together 
with uniform-disk fits.  
{\bf e:} Uniform-disk diameter (UDD) obtained by fitting the visibility 
points of all four data sets is plotted as a function of 
wavelength.  {\bf f:} Observed spectra of IRAS08002.  The absolutely 
calibrated MIDI spectrum 
is plotted with the filled circles, while the IRAS LRS is plotted with the 
open squares.
}
\label{vis_obs}
\end{figure*}

IRAS08002 was observed with MIDI on 2004 February 9, 10, and 11 
as part of the Science Demonstration Time program. 
A prism with a spectral resolution of $\lambda/\Delta \lambda \simeq 30$ 
at 10~\mbox{$\mu$m}\ 
was used to obtain spectrally dispersed fringes 
between 8 and 13~\mbox{$\mu$m}.  
A detailed description of the observing procedure is given in 
Przygodda et al. (\cite{przygodda03}), Leinert et al.\ (\cite{leinert04}), 
and Chesneau et al. (\cite{chesneau05a}, \cite{chesneau05b}).  
Four data sets were obtained using the 47~m baseline 
between the telescopes UT2 and UT3 (see Tables~\ref{table_obs} 
and \ref{table_calib} for the summary of the observations).  

We used two different MIDI data reduction packages: MIA developed 
at the Max-Planck-Institut f\"ur Astronomie and EWS developed 
at the Leiden Observatory.   While the MIA package is based on 
the power spectrum analysis, which measures the total power 
of the observed fringes (Leinert et al. \cite{leinert04}), 
the EWS software first corrects for optical path differences 
(instrumental as well as atmospheric delays) in each scan, and 
then, the fringes are coherently added (Jaffe \cite{jaffe04}).  

Figure~\ref{vis_obs} shows the calibrated visibilities of IRAS08002 
derived from the four data sets.  
In the figure, the means of the calibrated 
visibility values derived with the MIA and EWS packages are plotted 
(the results obtained with two reduction packages are found to be in 
good agreement for these data sets).  The data sets 
\#1 and \#3, which were obtained with approximately the same projected 
baseline lengths and position angles, show good agreement, 
indicating the good data quality and the reliability of the 
data reduction procedures.  The errors of the calibrated 
visibilities are typically $\pm 10$--15\%, and the error sources are
described in Ohnaka et al. (\cite{ohnaka05}).  
Figures~\ref{vis_obs}b--d show the 
observed visibilities plotted as a function of spatial frequency at 
three representative wavelengths between 8 and 13~\mbox{$\mu$m}, together 
with uniform-disk fits.  
We tentatively fit all the visibility points at a given 
wavelength with a uniform disk, regardless the position angles, 
because the limited position angle coverage of the currently 
available data 
does not allow us to draw a conclusion about the possible 
presence or absence of asymmetries.  
The resulting uniform-disk diameter is plotted as a function of 
wavelength in Fig.~\ref{vis_obs}e.  
It should be stressed, however, that 
we use uniform-disk fits to obtain some kind of representative 
angular size of the object and that 
the real intensity distribution of the object can be very 
different from a uniform disk.  

We also extracted the $N$-band spectrum of IRAS08002 from the MIDI 
data, as described in Ohnaka et al. (\cite{ohnaka05}).  
One of the calibrators, HD~67582 (K3III), for which an absolutely 
calibrated spectrum is available in Cohen et al. (\cite{cohen99}), 
was observed at similar air masses as IRAS08002 and 
used as a spectrophotometric standard star.  
The absolutely calibrated 
spectrum of IRAS08002 extracted from the MIDI data is shown in 
Fig.~\ref{vis_obs}f together with the IRAS LRS.  The MIDI spectrum 
shows good agreement with the IRAS LRS, confirming that the silicate 
emission spectrum has been stable for the last 20 years.  This is consistent 
with the result of the ISO observation by Yamamura et al. 
(\cite{yamamura00}), who found that the silicate emission features 
in another silicate carbon star V778~Cyg were stable 
for 14 years between the IRAS and ISO observations.  

Figure~\ref{vis_obs} reveals that 
all visibilities derived from the four data sets show a distinct 
wavelength dependence: a steady increase from 8 to $\sim$10~\mbox{$\mu$m}\ 
and a nearly constant part longward of 10~\mbox{$\mu$m}\ with a very 
slight decrease.  This wavelength 
dependence of the observed visibility translates into roughly constant 
uniform-disk diameters between 8 and 10~\mbox{$\mu$m}\ and a rather steep 
increase longward of 10~\mbox{$\mu$m}, as shown in Fig.~\ref{vis_obs}e.  
The wavelength dependence of the visibilities and uniform disk diameters 
observed in IRAS08002 resembles that observed for the Mira variable 
RR~Sco in the $N$ band presented in Ohnaka et al. (\cite{ohnaka05}).  
In the case of RR~Sco, Ohnaka et al. (\cite{ohnaka05}) interpret the 
constant uniform-disk diameters between 8 and 10~\mbox{$\mu$m}\ as due to 
the optically thick emission from warm, dense layers of 
H$_2$O and SiO gas extending to $\sim$2.3~\mbox{$R_{\star}$}, 
while the increase of the 
diameter longward of 10~\mbox{$\mu$m}\ can be attributed to an optically thin 
dust shell consisting of silicate and corundum.  
It should be stressed here, however, that this cannot be the case for 
IRAS08002, because the stellar angular diameter of IRAS08002 is estimated 
to be $\sim$1.7~mas if a linear radius of 400~\mbox{$R_{\sun}$}\ and a distance 
of 2.2~kpc (Engels \cite{engels94a}) are assumed.  
It is unlikely that those dense molecular layers extend to the 
region as far as $\sim$18--27~mas in radius ($\sim$21--32~\mbox{$R_{\star}$}) 
from the central star.  

The wavelength dependence of the visibility observed for IRAS08002 shows 
marked contrast to that observed for other objects with a prominent 
10~\mbox{$\mu$m}\ silicate feature.  Leinert et al. (\cite{leinert04}) 
observed a number of Herbig Ae/Be stars showing the amorphous and 
crystalline silicate emission features using MIDI.  
Most of the observed visibilities 
show a decrease from 8 to $\sim$9.5~\mbox{$\mu$m}\ and a gradual increase 
longward of 9.5~\mbox{$\mu$m}.  
The visibilities observed for the symbiotic Mira RX~Pup 
also exhibit the same trend (Driebe et al., in preparation).  
Obviously, the visibilities observed 
in IRAS08002 show the opposite shape compared to the visibilities 
observed in these objects.  

It should be mentioned here, however, that the current lack of 
sufficient complementary data on IRAS08002, such as high angular
resolution observations by speckle and/or aperture-masking
interferometry, high-resolution mid-infrared spectra, and polarimetric
observations, makes it difficult to obtain a unique picture of 
this object.  
We also note that no radiative transfer model for a 
circumbinary or circum-companion dust disk around silicate carbon stars 
has been presented in the literature.  
Therefore, we start from simple models with silicate dust alone, 
and after showing that such models 
have difficulties in reproducing the observations, 
we propose models with two grain species as possible scenarios.

\section{Dust radiative transfer code}
\label{sect_code}

\subsection{Basics of the Monte Carlo code}
\label{subsect_basics}

We have developed a Monte Carlo radiative transfer code (\mbox{\sf mcsim\_mpi}) 
and used it for the interpretation of the observations of IRAS08002.  
Our code \mbox{\sf mcsim\_mpi}\ can deal with arbitrary density distributions 
(spherically symmetric cases, axi\-symmetric cases, and general 
three-dimensional cases) as well as 
multiple grain species which may have different density distributions.  
Dust with multiple sizes can also be handled with \mbox{\sf mcsim\_mpi}, which 
simply treats dust grains of different sizes as different grain 
species.  
The calculation of polarization is also included in the code, and 
its application will be presented in Murakawa et al. (in 
preparation).  
The basics of the Monte Carlo technique are well explained somewhere else 
(e.g., Wolf et al. \cite{wolf99}; 
Niccolini et al. \cite{niccolini03}; Wolf \cite{wolf03}), 
and therefore, 
we only briefly describe the outline of the technique.  

In the Monte Carlo technique, a number of photon packets are emitted 
from the surface of the central star with a radius \mbox{$R_{\star}$}\ and an 
effective temperature \mbox{$T_{\rm eff}$}, 
and they travel through
the circumstellar dust envelope, interacting with dust grains.  
We divide the model space into many cells and assume the 
density and the temperature to be constant in each cell.  
The initial position of a photon packet on the stellar surface as well
as its direction is determined with a random number.  
The frequency of each initial photon packet is also chosen 
by a random number so that the spectral shape of emitted photon
packets 
should follow the blackbody radiation of the given effective temperature 
\mbox{$T_{\rm eff}$}.  The energy of each photon packet is given by 
$L_{\star} \Delta t / N$, 
where $L_{\star}$ is the stellar luminosity, $\Delta t$ is the 
time interval of the simulation, and $N$ is the total number of 
photon packets.
The optical depth ($\tau_{\nu}$) which every photon packet can proceed 
before the next interaction with dust grains (absorption or scattering) 
is chosen as $\tau_{\nu} = - \ln (1 - p)$, where $p$ is a 
random number uniformly distributed between 0 and 1.  As the photon 
packet moves along its path, we calculate the optical depth by 
\[
 \int \sum_{i=1}^{N_{\rm sp}} \rho_{i}( \mathbf{r} ) \, 
(\kappa_{\nu, i} + \sigma_{\nu, i}) \, d\ell, 
\]
where $\rho_{i} (\mathbf{r})$ is the number density of the $i$-th 
grain species at the position $\mathbf{r}$, $\kappa_{\nu, i}$ 
and $\sigma_{\nu, i}$ are the absorption and scattering cross sections 
of the $i$-th grain species, respectively, and $N_{\rm sp}$ is the 
number of different grain species.  The integration is 
performed along the path of the photon packet.  
The photon packet travels until this accumulated optical depth 
reaches $\tau_{\nu}$, the value chosen at the beginning.  

When the photon packet reaches the point of the next interaction with 
dust grains, it undergoes either absorption or scattering.  
In the presence of multiple grain species, the probability 
for absorption ($P_{\rm abs}$) or scattering 
($P_{\rm sca}$) with the $i$-th grain species is given by 
\[
 P_{\rm abs} = \frac{\rho_{i}(\mathbf{r}) \kappa_{\nu,i}}
 {\sum_{i=1}^{N_{\rm sp}} \rho_{i}(\mathbf{r}) 
 (\kappa_{\nu,i} + \sigma_{\nu,i})}, 
 P_{\rm sca} = \frac{\rho_{i}(\mathbf{r}) \sigma_{\nu,i}}
 {\sum_{i=1}^{N_{\rm sp}} \rho_{i}(\mathbf{r}) 
 (\kappa_{\nu,i} + \sigma_{\nu,i})},
\]
respectively.  In the case of scattering, the new direction of the 
photon packet is determined by a random number so that it samples 
the phase function of the grain with which the photon packet has 
just interacted.  While our code can deal with arbitrary phase 
functions, we assume isotropic scattering throughout the present
work for simplicity.  In the case of absorption, we adopt the temperature 
correction scheme developed by Bjorkman \& Wood (\cite{bjorkman01}).  
In their method, the energy of the photon packet just absorbed by a dust 
grain is used to calculate the dust temperature, and the 
photon packet is re-emitted immediately.  
The new direction is isotropically chosen, 
and the new frequency is chosen so that the re-emitted 
photon packet corrects the temperature of the spectrum previously 
emitted by the cell.  This scheme accelerates a Monte Carlo simulation
dramatically (provided that the opacity is independent of temperature).  
When the photon packet finally escapes the circumstellar envelope 
toward a given viewing angle, its final frequency is registered 
within $N_{\nu}$ frequency bins which are distributed 
equidistantly in the logarithmic scale between $\nu_{\rm min}$ and 
$\nu_{\rm max}$.  In the calculations presented here, we set $N_{\nu}$, 
$\nu_{\rm min}$, and $\nu_{\rm max}$ to be 256, $3 \times 10^{11}$~Hz 
(1000~\mbox{$\mu$m}), and $3 \times 10^{15}$~Hz (0.1~\mbox{$\mu$m}), respectively.  

Our code \mbox{\sf mcsim\_mpi}\ is parallelized with the use of 
LAM/MPI\footnote{http://www.lam-mpi.org}, which coordinates the 
code to run on different machines connected by the network.  
Specifically, a Monte 
Carlo simulation is started on each machine with a different 
random seed.  After the simulations on all the machines are completed, 
the results obtained on all the machines are collected, and 
the averages of the results such as temperature distributions, SEDs and 
images are calculated.  

After all photons have escaped the envelope, 
we store the monochromatic mean intensity ($J_{\nu}$) and the 
temperature in each cell 
so that images at a given wavelength viewed from an arbitrary angle 
can be created by ray tracing.  
Lucy (\cite{lucy99}) shows that 
the mean intensity in a particular cell is given by 
\[
 J_{\nu} \, d\nu = \frac{1}{4 \pi} \frac{1}{V} \frac{L_{\star}}{N} \sum_{d\nu} \ell, 
\]
where $V$ is the volume of the cell, 
and the summation is over all path lengths 
within the cell at issue for all photon packets with frequencies in 
$(\nu, \nu+d\nu)$.  With the monochromatic mean intensity and 
the temperature in each cell available, it is straightforward 
to calculate the intensity expected for any line of sight by ray tracing, 
which is expressed as 
\[
I_{\nu} = I_{\nu}^{\star} + \int S_{\nu} \, e^{-\tau_{\nu}} \, 
d\tau_{\nu}, \,\,
I_{\nu}^{\star} = B_{\nu}(\mbox{$T_{\rm eff}$}) \,\, e^{-\tau_{\nu}^{\star}}, 
\]
where $\tau_{\nu}^{\star}$ is the optical depth from the observer 
to the stellar surface along the line of sight defined by the viewing 
angle of the observer, and the integration is performed along this 
line of sight.  
The $I_{\nu}^{\star}$ term, which represents the intensity 
coming from the central star, must be added if the line of sight 
intersects with the central star.  Otherwise this term is set to 
zero. 
$S_{\nu}$ is the source function and, 
in the case of isotropic scattering, given by 
\[
S_{\nu} = 
\frac{\sum_{i=1}^{N_{\rm sp}}\rho_i(\mathbf{r})
(\kappa_{\nu,i} B_{\nu}(T_i(\mathbf{r})) + 
\sigma_{\nu,i} J_{\nu}(\mathbf{r}))}
{\sum_{i=1}^{N_{\rm sp}}\rho_{i}(\mathbf{r})
(\kappa_{\nu,i} + \sigma_{\nu,i})}, 
\]
where $T_i(\mathbf{r})$ is the temperature of the $i$-th grain species 
at the position $\mathbf{r}$, and $B_{\nu}(T_i(\mathbf{r}))$ is the 
Planck function of that temperature.  

Images thus produced using the 
ray-tracing technique are of much better S/N ratios compared to those 
created directly by collecting photons escaping the envelope during a 
Monte Carlo simulation.  
In the model calculations 
for IRAS08002, we use $\sim \!\! 10^6$--$5 \times 10^7$ photon 
packets, which produce SEDs as well as ray-tracing images with 
sufficient S/N ratios.

\subsection{Tests in spherically symmetric cases}
\label{subsect_1dtest}

\begin{figure}
\resizebox{\hsize}{!}{\rotatebox{0}{\includegraphics{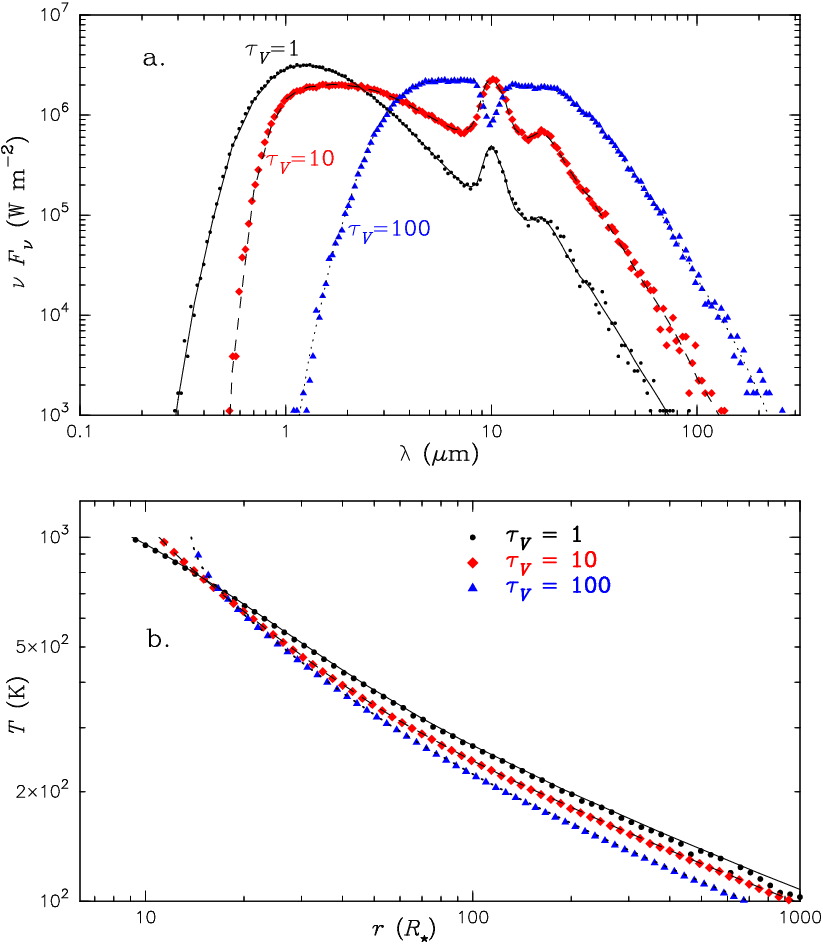}}}
\caption{{\bf a:} Comparison between the SEDs calculated with \mbox{\sf mcsim\_mpi}\ 
(see Sect.~\ref{subsect_basics}) 
and DUSTY for three different optical depths.   The effective 
temperature of the central star and the inner boundary temperature 
of the dust shell are set to be 3000~K and 1000~K, respectively.  
The results obtained with DUSTY are represented with the solid, 
dashed, and dotted lines, while the results obtained with \mbox{\sf mcsim\_mpi}\ 
are represented with the dots, filled diamonds, and filled 
triangles.  
{\bf b.} Comparison between the temperature distributions calculated 
with \mbox{\sf mcsim\_mpi}\ and DUSTY.  The meanings of the symbols are the same as 
in the panel {\bf a.}
}
\label{sed_1Dtest_2500}
\end{figure}

\begin{figure}
\resizebox{\hsize}{!}{\rotatebox{0}{\includegraphics{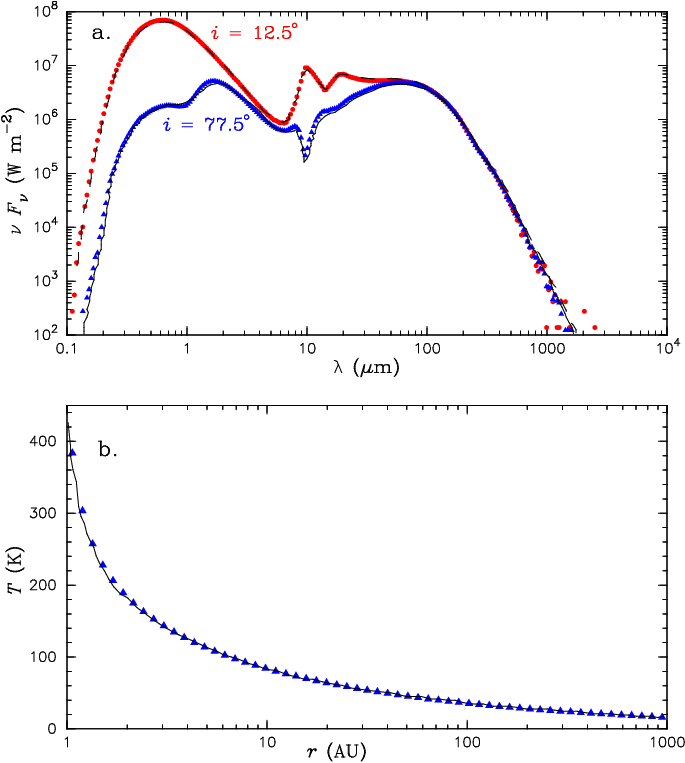}}}
\caption{Comparison between the results obtained with \mbox{\sf mcsim\_mpi}\ and 
the results of the benchmark tests for \mbox{$\tau_{V}$}\ = 100 
published in Pascucci et al. (\cite{pascucci04}).  
{\bf a.} Comparison of the SEDs calculated for two different 
inclination angles.  The results of the benchmark tests of Pascucci et al. 
(\cite{pascucci04}) are plotted with the solid and dashed lines, while 
the results obtained with \mbox{\sf mcsim\_mpi}\ are plotted with the filled circles 
and filled triangles.  
{\bf b.} Comparison of the temperature distributions obtained with 
\mbox{\sf mcsim\_mpi}\ and the benchmark tests.  The solid line represents the 
radial temperature distribution of the benchmark tests 
predicted for an angle of 2.5\degr\ measured 
from the equatorial plane.  The corresponding temperature distribution 
obtained with \mbox{\sf mcsim\_mpi}\ is represented with the filled triangles.  
}
\label{sed_2Dtest1}
\end{figure}

In order to check the reliability of the \mbox{\sf mcsim\_mpi}\ code, we performed test 
calculations in spherically symmetric cases and compared the 
results with those obtained with 
the DUSTY code, which is a publicly available radiative 
transfer program for spherically symmetric cases (Ivezi\'c \& Elitzur 
\cite{ivezic97}).  The spectrum of the central star is represented 
by blackbody radiation, and an effective temperature of 3000~K is assumed.  
The luminosity of the central star is set to $10^4$~\mbox{$L_{\sun}$}.  
The absorption and scattering 
cross sections are calculated from the complex refractive index 
of the warm silicate dust (Ossenkopf et al. \cite{ossenkopf92}) 
in the Mie theory for a single grain size of 0.1~\mbox{$\mu$m}, 
using the code of Bohren \& Huffman (\cite{bohren83}).  
Throughout the present work, we use 0.55~\mbox{$\mu$m}\ as a reference 
wavelength for 
the optical depth, and the optical depths ($\tau_V$) used for the 
test calculations are 1, 10, and 100.  
The dust density is assumed to be proportional to $r^{-2}$. 
The inner boundary of the dust shell (\mbox{$r_{\rm in}$}) is defined as a radius
where the dust temperature 
is equal to a pre-given dust sublimation temperature, for which we 
adopt 1000~K.  The outer boundary ($r_{\rm out}$) is set to be 
$10^3 \times \mbox{$r_{\rm in}$}$.  

For the calculations with \mbox{\sf mcsim\_mpi}, five 1.8~GHz Linux PCs are used, 
with $2 \times 10^6$ photon packets emitted on each machine. 
The dust envelope is divided into 100 cells 
in the radial direction, equidistantly on the logarithmic scale. 
We use the inner boundary radius derived with the DUSTY 
code as the input of our \mbox{\sf mcsim\_mpi}\ code.  
Figure~\ref{sed_1Dtest_2500} shows the SEDs and 
temperature distributions calculated with \mbox{\sf mcsim\_mpi}\ and DUSTY.  
Note that these SEDs are scaled to the 
values expected at the distance equal to the stellar radius 
defined as $(L_{\star}/4 \pi \sigma T_{\rm eff}^{4})^{1/2}$, 
where $\sigma$ is the Stefan-Boltzmann constant.    
The figure illustrates that the results 
obtained with both codes show good agreement in the three cases, 
demonstrating the reliability of the \mbox{\sf mcsim\_mpi}\ code.

\subsection{Test in axisymmetric cases}
\label{subsect_2dtest}

We further test our \mbox{\sf mcsim\_mpi}\ code in axisymmetric cases.  
Pascucci et al. (\cite{pascucci04}) performed benchmark test
calculations using five different multi-dimensional radiative 
transfer codes for a flattened, torus-like density distribution.  
We compare the result obtained with our \mbox{\sf mcsim\_mpi}\ 
code with the results of these benchmark tests using the same density 
distribution and input parameters.  
We calculate the model with $\tau_V$ = 100 in the equatorial direction, 
the highest optical depth in the models presented by 
Pascucci et al. (\cite{pascucci04}).  
Figure~\ref{sed_2Dtest1}a shows a 
comparison of SEDs viewed from two different inclination angles 
(the inclination angle $i$ is measured from the symmetry axis of the dust 
density distribution throughout the present work): 
$i = 12.5\degr$ (viewed nearly pole-on) and $i = 77.5\degr$ 
(viewed nearly edge-on).  These SEDs are scaled to the values 
expected at the distance of the stellar radius.  
 Figure~\ref{sed_2Dtest1}b shows the temperature distribution in 
the nearly equatorial direction.  
These figures 
demonstrate that the agreement between the results obtained with 
\mbox{\sf mcsim\_mpi}\ and the results of the benchmark 
tests of Pascucci et al. (\cite{pascucci04}) is good.

\section{Models with silicate dust alone}
\label{sect_modeling}

Using the Monte Carlo radiative transfer code described above, 
we calculate SEDs as well as $N$-band visibilities and compare with 
the observations of IRAS08002.  
The effective temperatures of IRAS08002 derived by previous 
authors are 2500~K (Kwok \& Chan \cite{kwok93}) and 3015~K 
(Bergeat et al. \cite{bergeat02}).  
In the present work, we assume the effective temperature to be 
2750~K, the mean value of this range. 
The luminosity of the central star is set to be $10^4$~\mbox{$L_{\sun}$}, 
assuming that the central 
star (the primary star with a carbon-rich photosphere) is the only 
heating source.  It would be reasonable to neglect the contribution of 
the putative companion, because its luminosity is considered to be
much lower than that of the primary star.  
For the dust opacity, 
we use the warm silicate dust of Ossenkopf et al. (\cite{ossenkopf92}) 
with a single grain size of 0.1~\mbox{$\mu$m}.  

\begin{figure}
\resizebox{\hsize}{!}{\rotatebox{0}{\includegraphics{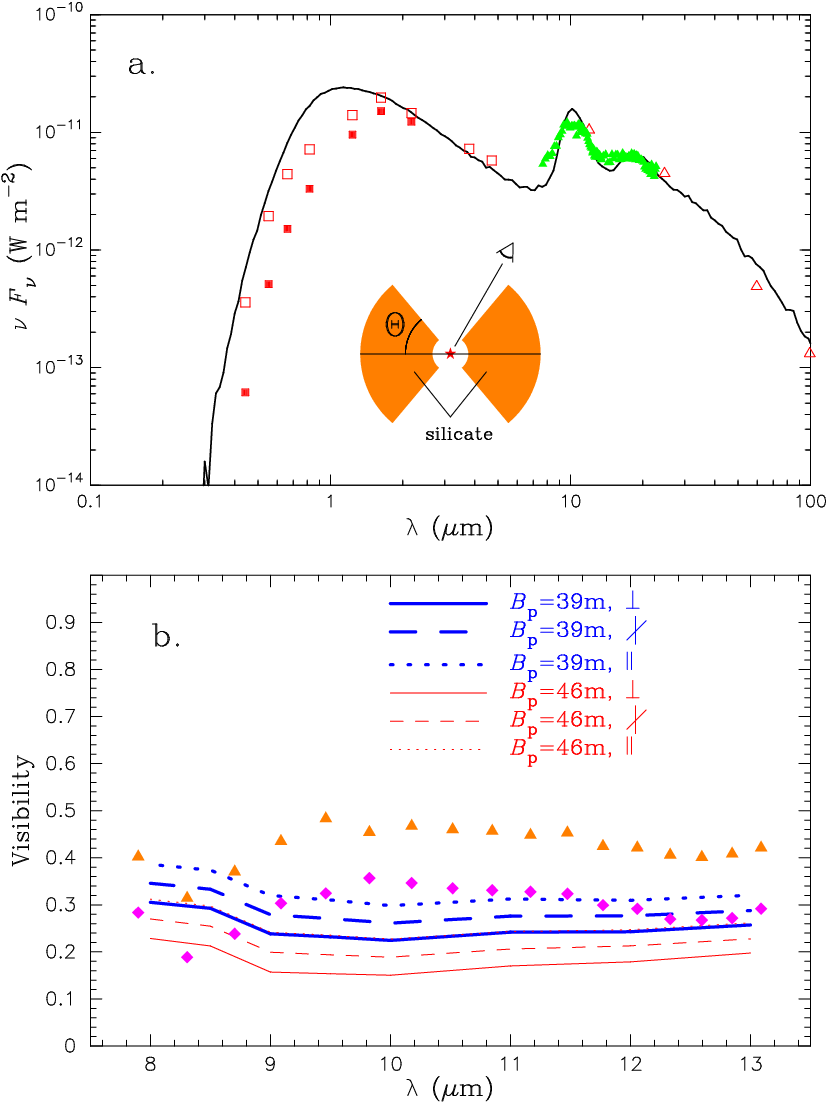}}}
\caption{
Axisymmetric disk model consisting of silicate dust alone.  
The parameters of the model are \mbox{$\tau_{V}^{\rm sil}$}\ = 15, 
\mbox{$r_{\rm in}$}\ = 15~\mbox{$R_{\star}$}\ (30~AU), $\Theta$ = 50\degr, and a radial 
density distribution proportional to $r^{-1.4}$.  
The model SED and visibilities are calculated for an inclination angle 
of 30\degr\ (measured from pole-on).  
{\bf a.} The observed SED is shown by the open squares 
(Le Bertre et al. \cite{lebertre90}, de-reddened with $A_V = 1.5$), 
open triangles (IRAS), and filled triangles (IRAS LRS).  
The original data of Le Bertre et al. (\cite{lebertre90}), 
not corrected for the interstellar extinction, are also plotted with 
the filled squares.  The model SED is represented with the solid line.  
{\bf b.} Filled triangles and diamonds: $N$-band visibilities measured 
with the 39~m and 46~m baselines (data sets \#2 and \#3), 
respectively.  
Thick and thin lines (solid, dashed, and dotted) represent the 
visibilities predicted for projected baseline lengths of 39 and 
46~m, respectively.  The solid, dashed, and dotted lines represent 
the model visibilities predicted for different position angles 
(solid line: perpendicular to the symmetry axis of the model 
intensity distribution, dotted line: parallel to the symmetry 
axis, dashed line: 45\degr\ with respect to the symmetry axis).  
See also Sect.~\ref{sect_modeling} and 
Fig.~\ref{bestmodel_silamc_same_image}.  
}
\label{model_2Dsil}
\end{figure}

First, we attempt to explain the observed SED and $N$-band 
visibilities using axisymmetric disk models (including spherical 
shell models as special cases) consisting of silicate dust alone.  
As depicted in the inset of Fig.~\ref{model_2Dsil}a, 
we represent this dust disk with a sphere from which a bipolar cavity 
is carved out.  Within the disk, the density 
depends only on the distance from the central star, while  
it is set to zero in the bipolar cavity.  
We split the disk into 100--200 cells in the radial direction and 
18--36 cells in the latitudinal direction within the disk.  
The free parameters are the inner 
boundary radius (\mbox{$r_{\rm in}$}), the optical depth in the equatorial direction 
(\mbox{$\tau_{V}^{\rm sil}$}), the disk half-opening angle measured from the 
equatorial plane ($\Theta$), 
and the exponent of the radial density distribution 
(density proportional to $r^{-p}$).  
The ranges of the changes of these parameters are \mbox{$r_{\rm in}$}\ = 
10 ... 25~\mbox{$R_{\star}$}\ with $\Delta \mbox{$r_{\rm in}$}$ = 5~\mbox{$R_{\star}$}, 
\mbox{$\tau_{V}^{\rm sil}$}\ = 5 ... 25 with $\Delta \mbox{$\tau_{V}^{\rm sil}$}$ = 5, 
$\Theta$ = 10\degr\ ... 90\degr\ with $\Delta \Theta = 20\degr$, 
and $p$ = 1.2 ... 2 with $\Delta p$ = 0.2.  
Note that a model with a disk half-opening angle ($\Theta$) of 
90\degr\ corresponds to a spherical shell model.  
The outer boundary radius is set to $500 \times \mbox{$r_{\rm in}$}$. 
This parameter has only minor effects on the 
SEDs in the wavelength range considered in the present work and 
$N$-band visibilities and, therefore, cannot be well constrained.  
The SEDs for 5--10 different inclination angles ($i$, as defined in 
Sect.~\ref{subsect_2dtest}) between 0\degr\ and 90\degr\ are calculated.  
For each model, we first compare the SEDs calculated for different 
inclination angles with the observed SED.  
For models which can reproduce the 
observed SED reasonably well, we produce images and visibilities 
in the $N$ band.

Figure~\ref{model_2Dsil}a shows an example of the models which 
can reproduce the observed SED rather well.  The parameters of this model 
are \mbox{$\tau_{V}^{\rm sil}$}\ = 15, \mbox{$r_{\rm in}$}\ = 15~\mbox{$R_{\star}$}, $p$ = 1.4, and 
$\Theta$ = 50\degr, and the SED plotted in Fig.~\ref{model_2Dsil}a was 
calculated for an inclination angle of 30\degr.  
The observed SED is constructed from the photometric data of 
Le Bertre et al. (\cite{lebertre90}) as well as the IRAS fluxes, 
together with the IRAS LRS.  
As for the interstellar extinction toward IRAS08002, 
Kwok \& Chan (\cite{kwok93}) derived $A_V$ = 1.5 from the SED 
fitting, while Bergeat et al. (\cite{bergeat02}) derived $A_J$ = 0.32, 
which translates into $A_V$ = 1.1 with $A_J = 0.87 E(B-V)$ and 
$A_V = 3.1 E(B-V)$ (Savage \& Mathis \cite{savage79}).  
These values are in agreement with the results 
obtained by Neckel \& Klare (\cite{neckel80}) for the estimated 
distance of IRAS08002 of 2.2~kpc.  
The photometric data of Le Bertre et al. (\cite{lebertre90}) are 
de-reddened using the method of Savage \& Mathis (\cite{savage79}) with 
$A_V$ = 1.5 (open squares in Fig.~\ref{model_2Dsil}).  We also plot the 
original data of Le Bertre et al. (\cite{lebertre90}) to show the 
effect of the interstellar extinction (filled squares).  
Note that the correction 
for the interstellar extinction is negligible at wavelengths longer 
than the $L$ band.  
The model flux is higher than the observed data at wavelengths 
shorter than $\sim$1~\mbox{$\mu$m}, but this may be attributed to 
the uncertainty of the effective temperature of the central star 
as well as the use of the blackbody for the stellar spectrum, 
neglecting the molecular absorption seen in usual carbon stars.  

It should be noted here that spherical shell models, which are 
special cases with $\Theta = 90\degr$, cannot reproduce the observed 
SED and $N$-band visibilities.  We found that the models, which can 
reproduce the SED from the near- to the far-infrared, predict 
the flux at wavelengths shorter than $\sim$1~\mbox{$\mu$m}\ to be 
considerably lower than the observations, that is, the central 
star appears to be too obscured.  This means that in the 
spherically symmetric geometry, the amount of silicate dust necessary 
to account for the observed infrared excess is not consistent with 
the low extinction toward the central star.  
As Lloyd-Evans (\cite{lloyd-evans90}) argues, this indicates that 
dust may exist in a disk, instead of a spherical shell.  
Although the agreement at optical wavelengths can be somewhat 
improved by adopting an effective temperature of 3000~K, 
it has turned out that such spherical models cannot reproduce 
the MIDI observations like the disk models discussed below.

The visibilities expected from the model shown in
Fig.~\ref{model_2Dsil}a are shown in Fig.~\ref{model_2Dsil}b as a 
function of wavelength.  
We need to estimate the angular radius of the central star to compare 
the predicted visibilities with the MIDI observations.  
We note that the model SEDs are scaled to fit the observed fluxes, and 
this scaling factor is given by $(\mbox{$R_{\star}$}/d)^{2}$, where $d$ is the 
distance to IRAS08002, and $\mbox{$R_{\star}$}/d$ is the angular radius of the 
central star.  
Using this angular radius, 
we can obtain model visibilities by calculating the two-dimensional 
Fourier transform of the intensity distribution of the object at 
a given wavelength, which is produced with the ray-tracing method 
described in Sect.~\ref{sect_code}.  
We also note here that since the model 
intensity distributions are not centrosymmetric except for models with 
$\Theta = 90\degr$, we need to 
specify the position angle of the symmetry axis of the disk relative 
to north to compare with the MIDI observations.  
This is equivalent to specifying the angle between the symmetry 
axis of the disk (projected onto the plane of the sky) and the 
projected baseline vector used in observations, and 
we use this angle as a free parameter. 
The dotted lines in Fig.~\ref{model_2Dsil}b represent the visibilities 
calculated with projected baseline vectors parallel to the symmetry 
axis of the disk in the plane of the sky, while the solid lines
represent those calculated with projected baseline vectors 
perpendicular to the symmetry axis.  
The dashed lines represent the visibilities calculated with 
projected baseline vectors at 45\degr\ with respect to the symmetry 
axis (see also Fig.~\ref{bestmodel_silamc_same_image} for 
the illustration of the orientation of these baseline vectors).  
In other words, for the cases plotted with the dotted, dashed, 
and solid lines, the position angles of the disk axis relative to north 
are given by P.A., P.A.$\pm$45\degr, and P.A.$\pm$90\degr, 
respectively, where 
P.A. is the position angle of a projected baseline used in our MIDI 
observations.  
The dotted and solid lines represent the full 
range of visibilities predicted for different position angles of the 
disk, and we check whether the observed visibilities can be reproduced 
within this range.  

Figure~\ref{model_2Dsil}b reveals that this disk model fails to 
reproduce the MIDI observations.  
The model predicts the visibilities to decrease from 8 to 10~\mbox{$\mu$m}, 
in clear disagreement with the MIDI observations.  
The optical depths of the disk models are larger at 10~\mbox{$\mu$m}\ 
than at 8~\mbox{$\mu$m}, due to the prominent silicate 
feature.   Also, the flux contribution of the central star is 
larger at 8~\mbox{$\mu$m}\ than at 10~\mbox{$\mu$m}.  
Therefore, if the spatial distribution of silicate dust 
is extended, as in the case of the models here, the angular size 
at 10~\mbox{$\mu$m}\ becomes remarkably larger than at 8~\mbox{$\mu$m}.  
A significant increase of the optical depth makes 
the decrease of the visibility from 8 to 10~\mbox{$\mu$m}\ less steep, but 
it cannot reproduce the observed increase, and the silicate emission 
spectrum also appears distorted.  
With a simple disk geometry used in the present work, 
we could not find a parameter set which can simultaneously reproduce 
the observed SED and the $N$-band visibilities.  
We also calculated disk models whose density distributions are 
proportional to $r^{-p} \times \exp (-(\theta/\Theta_{0})^{2})$, 
where $\Theta_0$ is a constant to define the geometrical thickness 
of the disk, and $\theta$ is the latitudinal angle measured from the 
equatorial plane.  However, the model visibilities show the same trend 
as found in the above models.  
As for the possible effects of more complicated disk geometries, 
we note here that Leinert et al. (\cite{leinert04}) present the $N$-band 
visibilities of Herbig Ae/Be stars (showing the 10~\mbox{$\mu$m}\ silicate 
feature) predicted by disk models 
with more realistic geometries such as the puffed-up rims and 
self-shadowing taken into account.  Still, these model visibilities are 
characterized by a steep decrease from 8 to 10~\mbox{$\mu$m}\ and a 
plateau with a slight increase up to 13~\mbox{$\mu$m}, which is obviously 
in disagreement with the MIDI observations of IRAS08002.  
Of course, the temperatures of the central stars of these Herbig 
Ae/Be stars as well as the physical properties of the disks are 
very different from those of IRAS08002.  
Therefore, 
while we cannot completely exclude the possibility that more 
complicated disk geometries may partially explain the MIDI observations 
of IRAS08002, 
and this possibility should be examined when more observational 
constraints become available, 
it seems unlikely that disk (and also spherical shell) models consisting 
of silicate dust alone can fully explain the observed SED and $N$-band 
visibilities.

\section{Models with two grain species}
\label{sect_multgrain}

As a possible scenario to explain the observed SED and $N$-band 
visibilities, we consider models with two grain species, silicate and 
some other grain species.  
Since the IRAS LRS and MIDI spectrum do not show prominent features 
other than the broad 10~\mbox{$\mu$m}\ feature, we tentatively consider 
the following three grain species, which exhibit no conspicuous 
spectral features in the wavelength range covered by these spectra. 

The first candidate is amorphous carbon.  
Given the fact that the central star is a carbon star, 
it is well possible that the circumstellar environment may have 
mixed chemistry consisting of oxygen-rich dust (silicate) and 
carbon-rich dust resulting from the current mass loss.  
The second candidate is large silicate grains.  
The opacity of a spherical grain with a radius $a$ 
becomes flat at wavelengths shorter than $\sim \!\! 2 \pi a$, 
showing little wavelength dependence.  
This means that large silicate grains 
($a \ga 3$~\mbox{$\mu$m}) do not show the 10~\mbox{$\mu$m}\ and 18~\mbox{$\mu$m}\ 
features and can be a candidate for the second grain species.  
There is also observational evidence for the presence of 
large grains in the circumstellar environment of silicate 
carbon stars, as discussed in Sect.~\ref{subsect_silsil}.  
The third candidate is metallic iron.  
The theoretical calculations 
of dust formation by Gail \& Sedlmayr (\cite{gail99}) 
suggest that metallic iron, which has a condensation 
temperature only by 50--100~K lower than that of silicate, 
can form in oxygen-rich AGB winds 
as separate grains or as inclusions within silicate grains.  
There is also observational evidence suggestive of the 
presence of metallic iron in the circumstellar envelope of 
cool, evolved stars (e.g., Harwit et al. \cite{harwit01}; 
Kemper et al. \cite{kemper02}; see also Sect.~\ref{subsect_silfe}).  

We note here that 
crystalline silicate was identified in the silicate 
carbon star IRAS09425-6040 (Molster et al. \cite{molster99}, \cite{molster01}). 
The slight hump at $\sim$11.5~\mbox{$\mu$m}\ 
seen in the spectra of IRAS08002 coincides with the position of 
a crystalline silicate feature.  
However, since IRAS08002 was not observed with ISO, and 
the IRAS LRS as well as the MIDI spectrum is of low spectral
resolution, it is not possible to confirm the presence of crystalline 
silicate unambiguously.  
Thus, while we stress the importance of high-resolution mid-infrared 
spectroscopy to confirm the presence or absence of crystalline 
silicate toward IRAS08002, we do not consider models with crystalline 
silicate in the present work.

\subsection{Models with silicate and amorphous carbon}
\label{subsect_silamc}

\begin{figure*}
\sidecaption
\includegraphics[width=12cm]{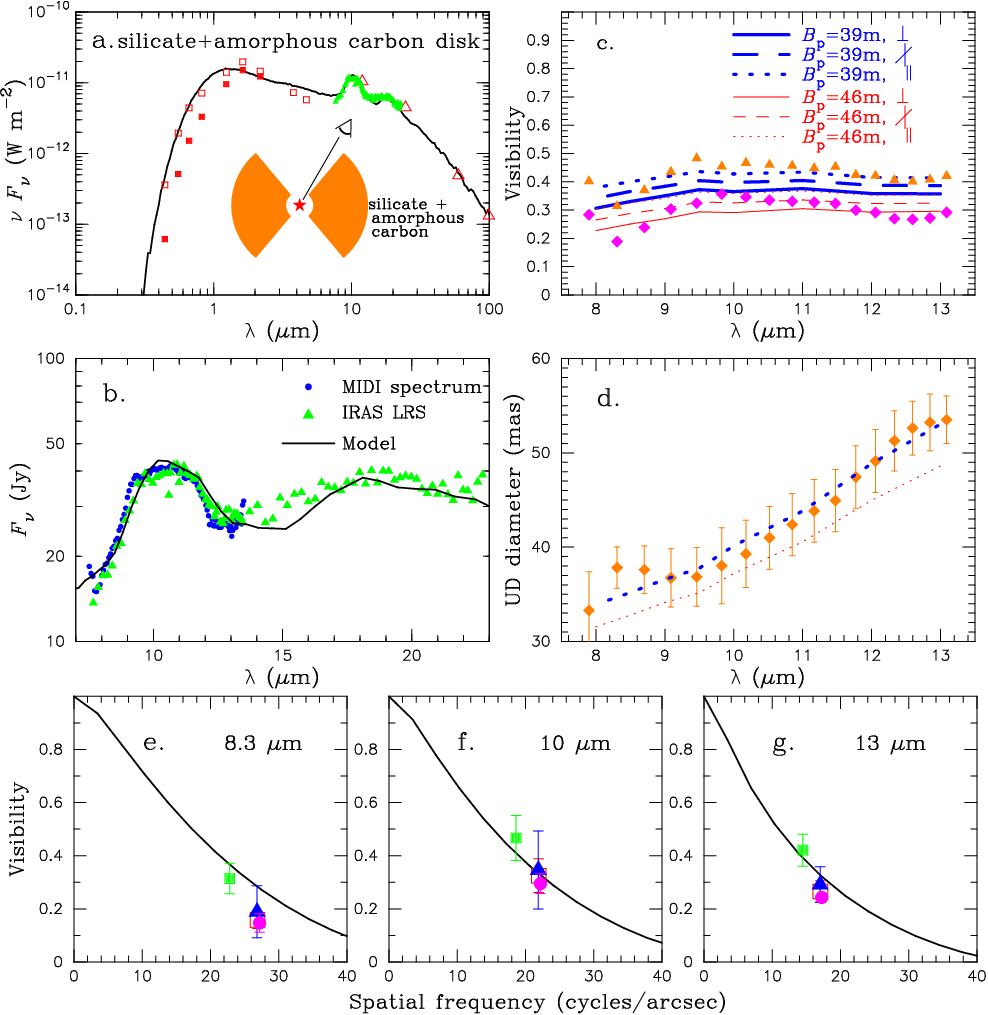}
\caption{The best-fit disk model consisting of silicate and amorphous 
carbon (see Sect.~\ref{subsect_silamc}).  
The parameters of the model are \mbox{$\tau_{V}^{\rm sil}$}\ = 25, \mbox{$\tau_{V}^{\rm amc}$}\ = 15, 
\mbox{$r_{\rm in}$}\ = 15~\mbox{$R_{\star}$}\ (30~AU), \mbox{$p$}\ = 1.6, and $\Theta$ = 50\degr.  
The model SED and visibilities are calculated for an inclination angle 
of 30\degr.  
See also the legend to Fig.~\ref{model_2Dsil} for the references of the 
symbols.  
{\bf a.} 
Comparison between the observed and model SEDs.  
{\bf b.} 
Comparison between the observed silicate emission spectra 
and the predicted spectrum. 
{\bf c.} Filled triangles and diamonds: $N$-band visibilities measured 
with the 39~m and 46~m baselines, respectively.  
Thick and thin lines (solid, dashed, and dotted) represent the 
corresponding model visibilities.  See the legend to Fig.~\ref{model_2Dsil} 
for the meanings of the different line types.  
{\bf d.} The observed uniform-disk diameters are plotted 
with the filled diamonds.  
The thick and thin dotted lines represent the uniform-disk diameters 
predicted for the 39~m and 46~m baselines, respectively, with 
the baseline vector parallel to the symmetry axis.  
{\bf e.--g.} The model visibilities predicted with the baseline vector 
parallel to the symmetry axis are plotted (solid lines) together 
with the observed visibilities.  See Fig.~\ref{vis_obs} for 
the references of the symbols. 
}
\label{bestmodel_silamc_same}
\end{figure*}

Firstly, we examine models containing silicate and amorphous carbon.  
At the moment, 
we assume that the two grain species coexist in an axisymmetric 
disk, as depicted in the inset of Fig.~\ref{bestmodel_silamc_same}a 
(another possible geometry is discussed later). 
The free parameters in this geometry are the same as described in 
Sect.~\ref{sect_modeling}, except that we now have an additional 
parameter to specify the optical depth of amorphous carbon 
in the radial direction (\mbox{$\tau_{V}^{\rm amc}$}).  
The density distributions of silicate and amorphous carbon grains are 
assumed to be proportional to $r^{-p}$.  
The warm silicate dust of Ossenkopf et al. (\cite{ossenkopf92}) 
and the amorphous carbon of Rouleau \& Martin (\cite{rouleau91}) 
are used, with a grain size of 0.1~\mbox{$\mu$m}\ adopted for both 
grain species.  
The ranges of the changes of the parameters are as follows: 
\mbox{$\tau_{V}^{\rm sil}$}\ = 15 ... 35 with $\Delta \mbox{$\tau_{V}^{\rm sil}$} = 5$, 
\mbox{$\tau_{V}^{\rm amc}$}\ = 5 ... 25 with $\Delta \mbox{$\tau_{V}^{\rm amc}$} = 5$,  
\mbox{$r_{\rm in}$}\ = 10 ... 25~\mbox{$R_{\star}$}\ with $\Delta \mbox{$r_{\rm in}$} = 5$~\mbox{$R_{\star}$}, 
$\Theta$ = 10\degr\ ... 90\degr\ with $\Delta \Theta = 20\degr$, 
and $p$ = 1.4, 1.6, 1.8, 2.0.  
The outer boundary radius is fixed to $500 \times 
\mbox{$r_{\rm in}$}$, because the effects of this parameter on SEDs and 
$N$-band visibilities are minor, and therefore, it cannot be 
well constrained by the observational data available at the 
present.

Figure~\ref{bestmodel_silamc_same} shows the results obtained 
by the best-fit model in this geometry.  
The disk inner radius is found to be 15~\mbox{$R_{\star}$}, while 
the exponent of the density distribution of silicate and amorphous 
carbon dust is derived to be 1.6.  
The optical depths of silicate and amorphous carbon dust 
are found to be 25 and 15 at 0.55~\mbox{$\mu$m}\ 
(1.9 and 0.1 at 10~\mbox{$\mu$m}), respectively.  
The disk half-opening angle and the inclination angle are found 
to be 50\degr\ and 30\degr, respectively.  
We estimate the uncertainties of these parameters by varying them 
by a small amount around the parameter set of the best-fit model.  
The uncertainties of the optical depths and the inner boundary 
radius are estimated to be approximately 20\%.   
The uncertainty of \mbox{$p$}\ is estimated to be roughly $\pm 0.1$.  
We estimate the uncertainty of the disk half-opening 
angle and that of the inclination angle to be roughly $\pm 10$\degr.  

Figures~\ref{bestmodel_silamc_same}a and \ref{bestmodel_silamc_same}b 
show that the model SED is in fair agreement with the observed one 
in the wavelength range from the optical to the far-infrared. 
Figures~\ref{bestmodel_silamc_same}c and 
\ref{bestmodel_silamc_same}e--g show a comparison between the 
observed $N$-band visibilities and those predicted by this model.  
Given the errors of the observed visibilities (see 
Figs.~\ref{bestmodel_silamc_same}e--g), 
the visibilities predicted for the baseline vector 
with 45\degr\ or parallel with respect to the symmetry axis (dashed 
and dotted lines in Fig.~\ref{bestmodel_silamc_same}c, respectively) 
can approximately reproduce the 
observed $N$-band visibilities, although the predicted visibilities 
tend to be flatter than the observations between 8 and 10~\mbox{$\mu$m}\ 
(this point will be discussed in Sect.~\ref{sect_discuss}).  
This rough agreement can also be seen in 
Fig.~\ref{bestmodel_silamc_same}d, where a comparison between the 
observed and predicted uniform-disk diameters is shown.  

\begin{figure*}
\sidecaption
\includegraphics[width=12cm]{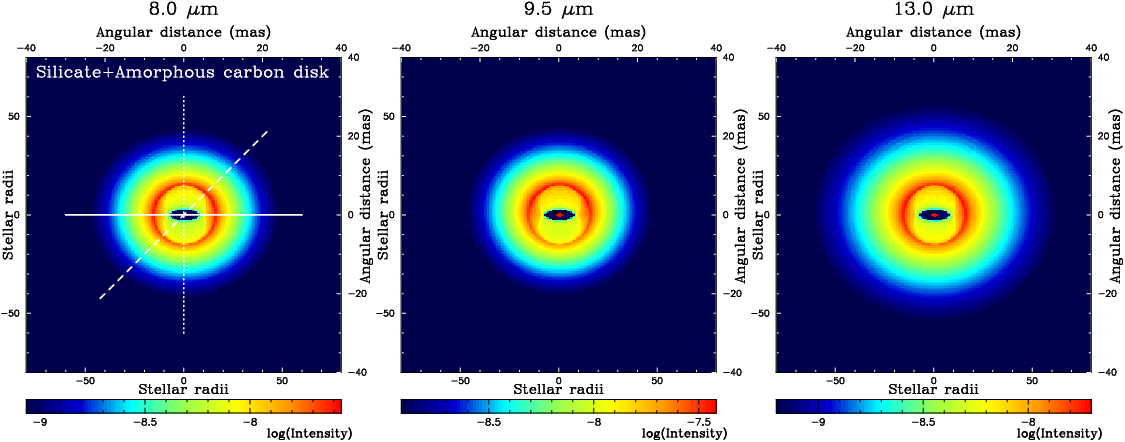}
\caption{The mid-infrared images predicted by the best-fit model 
containing silicate and amorphous carbon dust.  In each image, the 
color scale is normalized with the maximum intensity (excluding 
the central star) and the minimum intensity set to $3 \times 10^{-2} 
\times $~maximum intensity (the color scale on the central star is 
saturated).  The model parameters are 
given in the legend to Fig.~\ref{bestmodel_silamc_same} as well as 
in Sect.~\ref{subsect_silamc}.  
The solid, dashed, and dotted lines in the left panel illustrate the 
orientation of the baseline vectors used in the visibility
calculations.  
}
\label{bestmodel_silamc_same_image}
\end{figure*}

The reason this model with two grain species can fairly reproduce the 
observed wavelength dependence of the $N$-band visibilities can be 
explained as follows.  We first note that the model intensity 
distribution of the amorphous carbon component is more extended than 
that of the silicate component, because the high optical depth 
of this latter component makes the temperature gradient in the 
radial direction steeper than that of the former component.  
The total visibility is the weighted sum of the visibilities of 
these two components.  The weights of both components 
are given by the ratio of the fractional fluxes of the two components, 
which varies significantly between 8 and 13~\mbox{$\mu$m}\ due to the 
10~\mbox{$\mu$m}\ silicate feature.  Combination of this variation of the 
flux contributions of both components and their different 
angular sizes (and also different wavelength dependences of the
angular sizes) results in the $N$-band visibilities which are very 
different from those expected from disk models with silicate dust 
alone, and that can approximately reproduce the MIDI observations.  
Figure~\ref{bestmodel_silamc_same_image} shows the images 
at 8, 9.5, and 13~\mbox{$\mu$m}\ predicted by the best-fit 
model with silicate and amorphous carbon.  
The images at 8 and 9.5~\mbox{$\mu$m}\ are of approximately the 
same size, while the image at 13~\mbox{$\mu$m}\ appears more extended 
than those at 8 and 9.5~\mbox{$\mu$m}, corresponding to 
the wavelength dependence of the observed uniform-disk 
diameters.

The temperatures of silicate and amorphous carbon dust at the 
inner boundary of the disk are approximately 730 and 900~K, 
respectively. The temperature of silicate dust is remarkably lower 
than condensation temperatures of silicate of $\sim$1000~K derived for 
AGB stars with high mass-loss rates 
($\dot{M} \ga 5 \times 10^{-5}$~$M_{\sun}$~yr$^{-1}$), 
but in agreement with the inner boundary temperatures found for 
objects with lower mass-loss rates (e.g., Lorenz-Martins \& Pompeia 
\cite{lorenz-martins00}; Suh \cite{suh04}).  
The temperature of amorphous carbon found here 
is also lower than most of the values ($\sim$1500~K) derived for carbon 
stars by Groenewegen (\cite{groenewegen95}) and Groenewegen et al. 
(\cite{groenewegen98}), but not exceptionally low.  
It should be noted, however, that the temperature at the disk inner
boundary may not necessarily correspond to the condensation 
temperatures of these grain species, because dust 
formation in such a disk as considered here can be 
different from that expected in mass outflows in usual AGB stars.  

The total dust mass in the above model 
is $3.0 \times 10^{-3}$~\mbox{$M_{\sun}$}\ (silicate) and $6.6 \times 10^{-4}$~\mbox{$M_{\sun}$}\ 
(amorphous carbon), with bulk densities of 3.7~g~cm$^{-3}$ 
and 2.3~g~cm$^{-3}$ assumed for silicate and amorphous carbon, 
respectively.  
This translates into a total disk mass of 0.4~\mbox{$M_{\sun}$}\ with 
a gas-to-dust-ratio of 100.  However, the estimated disk mass depends 
on the assumed outer boundary radius.  With an outer boundary radius 
of $10^{3} \times \mbox{$r_{\rm in}$}$, which can also provide reasonable 
agreement with the observations, the disk mass increases to 0.9~\mbox{$M_{\sun}$}.  
If we assume a mass loss rate of the order of 
$10^{-6}$--$10^{-5}$~\mbox{$M_{\sun}$}~yr$^{-1}$ (dust mass loss rate 
$\sim \!\! 10^{-8}$--$10^{-7}$~\mbox{$M_{\sun}$}~yr$^{-1}$) 
for the primary star in its oxygen-rich phase, 
the above mass of silicate dust can be 
accumulated in $\sim \! \! 10^{5}$~yr, if all material shed by 
mass loss is stored in the disk.  
While the lifetime of the circumbinary disk is uncertain, 
this time scale of mass loss seems to be plausible in terms of 
the time scale of stellar evolution at the AGB.  

One concern with this model is the rather high 
optical depth of amorphous carbon dust ($\mbox{$\tau_{V}^{\rm amc}$} \simeq 15$) 
and its connection to the current mass loss rate of the 
central star, which is now a carbon-rich AGB star.  
While such a high optical depth is not uncommon among 
dust-enshrouded carbon stars (e.g., Groenewegen et al. \cite{groenewegen98}; 
Lorenz-Martins et al. \cite{lorenz-martins01}), the mass loss rates found 
in those objects are also rather high in general ($\dot{M} \ga 
10^{-5}$~\mbox{$M_{\sun}$}~yr$^{-1}$), and the central stars 
are heavily obscured by thick dust shells.  
However, the observed SED of IRAS08002 does not suggest that 
the central star is significantly obscured.  
Therefore, if the current mass 
loss rate of IRAS08002 is of the order of $10^{-5}$~\mbox{$M_{\sun}$ yr$^{-1}$}, 
the mass loss may not be spherically symmetric.  
This means that the current (carbon-rich) mass loss 
may be taking place preferably toward 
the disk, which was presumably formed by oxygen-rich material 
previously shed by the primary star.  
However, the mass loss phenomenon in binary systems, in particular 
with a (circumbinary) disk, is not well understood, and it still 
remains to be studied whether the mass outflow can be directed 
toward the disk.  

We also calculated models where silicate grains are 
present in such a disk as considered above but 
amorphous carbon is distributed in a spherical shell 
resulting from the current mass loss of the central carbon star.  
In these models, the inner boundary radii for both grain 
species are set to be equal for simplicity.  
However, we could not find a parameter 
set that can provide reasonable agreement with the observed 
SED and $N$-band visibilities.  This is because the high optical 
depths of the amorphous carbon shell lead to significant 
obscuration of the central star, which is in disagreement 
with the observed SED of IRAS08002 as discussed above.  
If we decrease the optical depth 
of the amorphous carbon shell to match the observed SED, 
the model then approaches the disk model consisting only of silicate 
dust described in Sect.~\ref{sect_modeling} and fails to reproduce 
the MIDI observations.  
However, since we cannot entirely exclude the 
possibility that more complicated disk geometries together with 
the amorphous carbon shell as well as addition of another 
grain species might explain the observations, 
more observational data are needed to test such complicated 
models.  

\begin{figure*}
\sidecaption
\includegraphics[width=12cm]{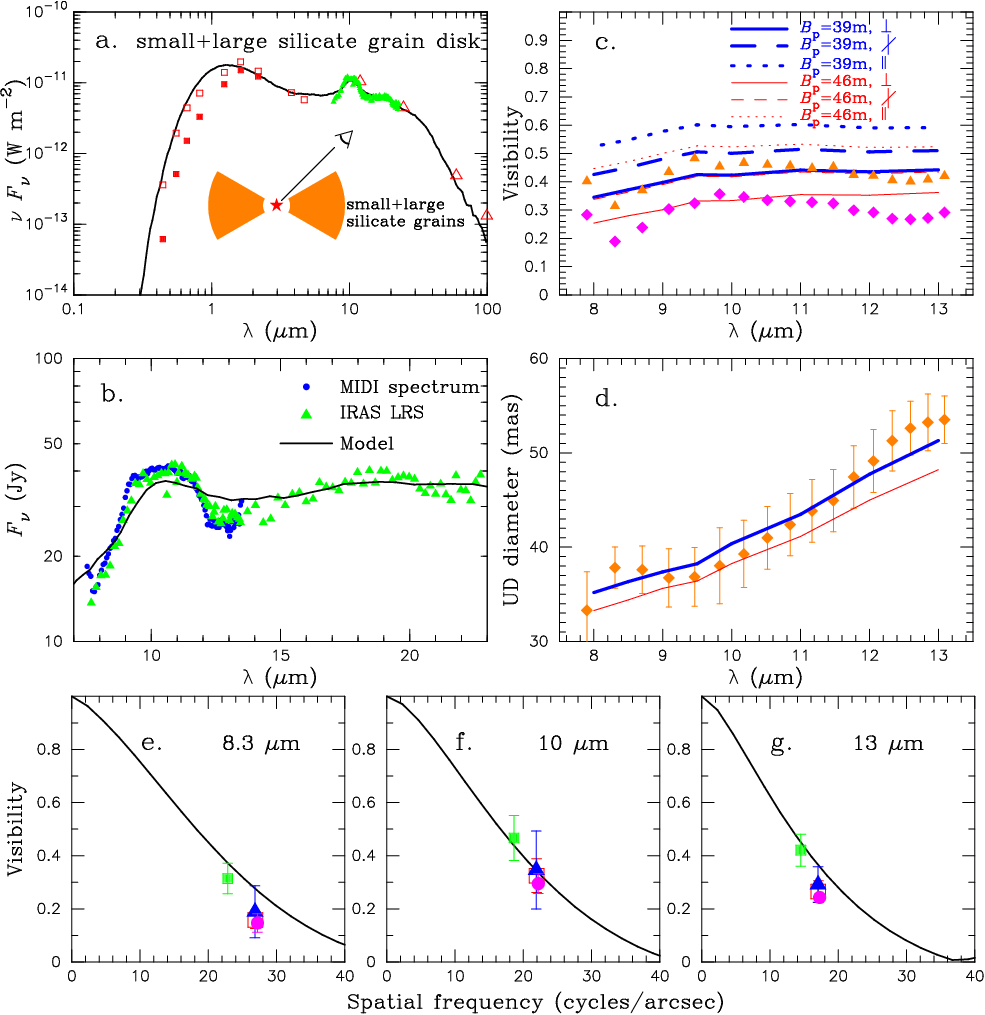}
\caption{
The best-fit disk model consisting of small and large silicate grains 
(see Sect.~\ref{subsect_silsil}).  
The parameters of the model are \mbox{$\tau_{V}^{\rm sil}$}\ = 25, \mbox{$\tau_{V}^{\rm SIL}$}\ = 2.0, 
\mbox{$r_{\rm in}$}\ = 15~\mbox{$R_{\star}$}\ (30~AU), \mbox{$p$}\ = 1.8, 
and $\Theta$ = 30\degr.  
The model SED and visibilities are calculated for an inclination angle 
of 45\degr.  
See also the legend to Fig.~\ref{model_2Dsil} for the references of the 
symbols. 
{\bf a.} 
Comparison between the observed and model SEDs.  
{\bf b.} 
Comparison between the observed silicate emission spectra 
and the predicted spectrum. 
{\bf c.} Filled triangles and diamonds: $N$-band visibilities measured 
with the 39~m and 46~m baselines, respectively.  
Thick and thin lines (solid, dashed, and dotted) represent the 
corresponding visibilities.  See the legend to Fig.~\ref{model_2Dsil} 
for the meanings of the different line types.  
{\bf d.} The observed uniform-disk diameters are plotted 
with the filled diamonds.  
The thick and thin solid lines represent the uniform-disk diameters 
predicted for the 39~m and 46~m baselines, respectively, with 
the baseline vector perpendicular to the symmetry axis.  
{\bf e.--g.} The model visibilities predicted with the baseline vector 
perpendicular to the symmetry axis are plotted (solid lines) together 
with the observed visibilities.  See Fig.~\ref{vis_obs} for 
the references of the symbols. 
}
\label{bestmodel_silsil_same}
\end{figure*}

\subsection{Models with small and large silicate grains}
\label{subsect_silsil}

\begin{figure*}
\sidecaption
\includegraphics[width=12cm]{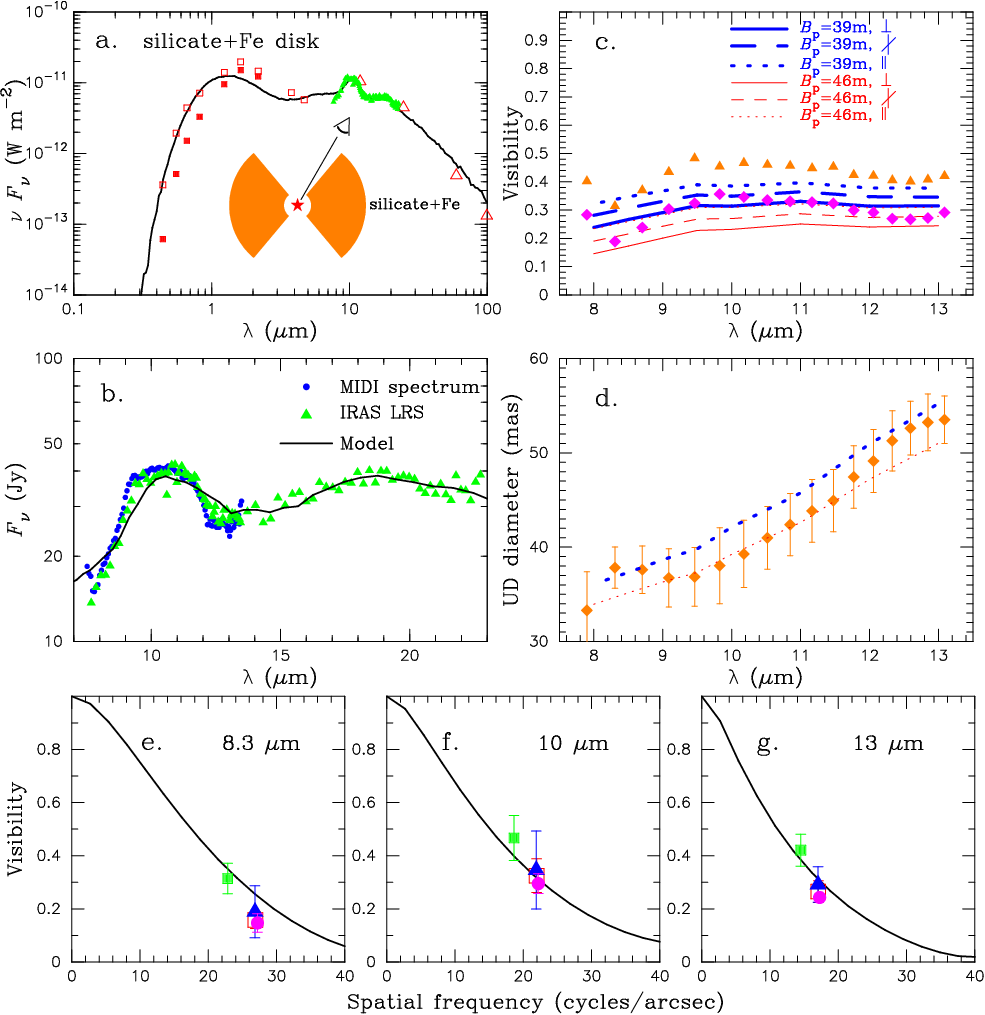}
\caption{
The best-fit disk model consisting of silicate and metallic iron 
grains (see Sect.~\ref{subsect_silfe}).  
The parameters of the model are \mbox{$\tau_{V}^{\rm sil}$}\ = 20, 
\mbox{$\tau_{V}^{\rm Fe}$}\ = 3.0, \mbox{$r_{\rm in}$}\ = 20~\mbox{$R_{\star}$}\ (40~AU), \mbox{$p$}\ = 1.6, 
and $\Theta$ = 50\degr.  
The model SED and visibilities are calculated for an inclination angle 
of 30\degr.  
See also the legend to Fig.~\ref{model_2Dsil} for the references of the 
symbols. 
{\bf a.} 
Comparison between the observed and model SEDs.  
{\bf b.} 
Comparison between the observed silicate emission spectra 
and the predicted spectrum. 
{\bf c.} Filled triangles and diamonds: $N$-band visibilities measured 
with the 39~m and 46~m baselines, respectively.  
Thick and thin lines (solid, dashed, and dotted) represent the 
corresponding model visibilities.  See the legend to Fig.~\ref{model_2Dsil} 
for the meanings of the different line types.  
{\bf d.} The observed uniform-disk diameters are plotted 
with the filled diamonds.  
The thick and thin dotted lines represent the uniform-disk diameters 
predicted for the 39~m and 46~m baselines, respectively, with 
the baseline vector parallel to the symmetry axis.  
{\bf e.--g.} The model visibilities predicted with the baseline vector 
parallel to the symmetry axis are plotted (solid lines) together 
with the observed visibilities.  See Fig.~\ref{vis_obs} for 
the references of the symbols. 
}
\label{bestmodel_silfe_same}
\end{figure*}

Secondly, we present models consisting of small and large silicate 
grains.  The spatial distributions of the two grain species are the 
same as used in the previous subsection, except that large silicate 
grains, instead of amorphous carbon, coexist with small silicate 
grains in the disk.  The spatial distributions of small and large 
silicate grains may not entirely overlap due to dust formation and 
growth processes, but we adopt this geometry for simplicity in the 
present work.  
We use the opacity of the warm silicate 
(Ossenkopf et al. \cite{ossenkopf92}) and tentatively adopt a grain 
size of 5~\mbox{$\mu$m}\ for large silicate grains and 
0.1~\mbox{$\mu$m}\ for small silicate grains.  
The optical depths of small and large silicate grains are 
denoted as \mbox{$\tau_{V}^{\rm sil}$}\ and \mbox{$\tau_{V}^{\rm SIL}$}, respectively, and 
the exponents of the radial density distributions 
of these two grain species are set to be equal 
and denoted as \mbox{$p$}.  
The ranges of the changes of 
these parameters are as follows: \mbox{$\tau_{V}^{\rm sil}$}\ = 15 ... 35 with 
$\Delta \mbox{$\tau_{V}^{\rm sil}$} = 5$, \mbox{$\tau_{V}^{\rm SIL}$}\ = 1 ... 3 with 
$\Delta \mbox{$\tau_{V}^{\rm SIL}$} = 0.5$, 
\mbox{$r_{\rm in}$}\ = 10 ... 25~\mbox{$R_{\star}$}\ with $\Delta \mbox{$r_{\rm in}$} = 5$~\mbox{$R_{\star}$}, 
$\Theta$ = 10\degr\ ... 90\degr\ with $\Delta \Theta = 20\degr$, 
and \mbox{$p$}\ = 1.4, 1.6, 1.8, 2.0.  
The outer radius of the disk is set to $500 \times \mbox{$r_{\rm in}$}$.

Figure~\ref{bestmodel_silsil_same} shows the best-fit model 
consisting of small and large silicate grains.  
The optical depths of small and large silicate grains 
are $25 \pm 5$ and $2.0 \pm 0.5$ , respectively.  
These optical depths correspond to 1.9 and 2.5 at 10~\mbox{$\mu$m}, 
respectively.  
The disk has an inner boundary radius of $15 \pm 3$~\mbox{$R_{\star}$}\ and 
a half-opening angle of $30\degr \pm 10\degr$ with $p = 1.8 \pm 0.1$.  
The inclination angle is derived to be $45\degr \pm 10\degr$.  
Figures~\ref{bestmodel_silsil_same}a and \ref{bestmodel_silsil_same}b 
illustrate that this model can approximately reproduce 
the whole SED as well as the silicate emission spectrum, 
although the predicted silicate emission feature appears 
too broad compared to the observed spectra.  
A decrease of the optical depth of large silicate grains can provide 
a better match to the observed silicate spectra, but the $N$-band 
visibilities predicted by such models do not show the 
increase from 8 to 10~\mbox{$\mu$m}\ as observed.  
Figures~\ref{bestmodel_silsil_same}c and \ref{bestmodel_silsil_same}e--g 
demonstrate that 
the $N$-band visibilities predicted for the 
baseline vector perpendicular to the symmetry axis of the model 
intensity distribution 
are in rough agreement with the MIDI observations 
(thin and thick solid lines), 
although the model visibilities are still somewhat too flat 
compared to those observed.  
It should be noted here that the intensity distribution of the 
large silicate component is more extended than that of the 
small silicate component, because the former component shows 
much more scattered light even in the 10~\mbox{$\mu$m}\ region 
due to the large grain size.  As discussed above, 
the variation of the flux ratio of these two components 
with different angular sizes (and different wavelength 
dependences of the angular sizes) 
produces visibilities which show a wavelength 
dependence roughly in agreement with the MIDI observations.  

The temperatures of the small and large silicate grains at the 
disk inner boundary are found to be $\sim$750~K and $\sim$590~K, 
respectively.  The mass of each component is 
$8.5 \times 10^{-4}$~\mbox{$M_{\sun}$}\ (small silicate grains) and 
$1.8 \times 10^{-3}$~\mbox{$M_{\sun}$}\ (large silicate grains).  
With a gas-to-dust ratio of 100, the total disk mass is derived 
to be 0.3~\mbox{$M_{\sun}$}.  This disk mass is comparable to that derived for 
the silicate+amorphous carbon model and will be discussed 
in Sect.~\ref{sect_discuss}.  

It is worth noting that 
there is growing observational evidence for the presence of large grains 
in mass-losing, evolved stars.  Jura et al. (\cite{jura01}) conclude 
that the far-infrared and sub-mm fluxes observed toward the silicate 
carbon star BM~Gem can be explained by grains as large as 0.1~mm.  
Molster et al. (\cite{molster01}) consider large grains 
($\sim$2~\mbox{$\mu$m}\ and $\sim$300~\mbox{$\mu$m}) to explain the mid-infrared 
spectrum of the silicate carbon star IRAS09425-6040.  
The lack of the 10~\mbox{$\mu$m}\ silicate feature in this object 
may even suggest the absence of grains smaller than $\sim$2~\mbox{$\mu$m}.  
The time scale for grain growth by grain-grain collisions 
(e.g., Jura \& Kahane \cite{jura99}; 
Yamamura et al. \cite{yamamura00}) implies that such large 
grains can form by coagulation, provided that the disk is stable 
(see Sect.~\ref{sect_discuss} for discussion on the stability 
of the disk around silicate carbon stars).  
The presence of large grains is also inferred in evolved stars 
other than silicate carbon stars.  For example, 
Jura et al. (\cite{jura97}) surmise that large grains may be 
responsible for the fluxes at radio wavelengths observed for 
the post-AGB star Red Rectangle.  The radiative transfer modeling 
of Men'shchikov et al. (\cite{menshchikov02}) for this object 
also lends support to the existence of grains as large as 
0.2~cm.  
In order to confirm the presence or absence of large grains in 
IRAS08002, far-infrared and sub-mm observations would be crucial.

\subsection{Models with silicate and metallic iron grains}
\label{subsect_silfe}

As the third possible scenario, we present models consisting of 
silicate and metallic iron grains.  The model geometry 
is the same as in the previous subsections (see also the inset of 
Fig.~\ref{bestmodel_silfe_same}a).  The opacity 
of silicate is calculated from the complex refractive index 
of the warm silicate of Ossenkopf et al. (\cite{ossenkopf92}) 
for a spherical grain with $a = 0.1$~\mbox{$\mu$m}.  
We note that Kemper et al. (\cite{kemper02}) 
found a satisfactory fit to the observed SED of the OH/IR star 
OH127.8+0.0 by including the opacity of non-spherical metallic iron 
grains, which were represented with a continuous distribution of 
ellipsoids (CDE, Bohren \& Huffman \cite{bohren83}).  
Kemper et al. (\cite{kemper02}) demonstrate that the 
opacities of metallic iron calculated in the Mie theory for a 
spherical grain and with CDE show a substantial difference.  
They also found that the amount of iron obtained from their 
modeling with non-spherical iron grains is consistent with 
the average interstellar abundances of Fe and Si, while 
the use of spherical iron grains results in a noticeable 
overabundance of Fe.  
Therefore, for metallic iron, we use the 
opacity calculated with CDE using the complex refractive 
index of Ordal et al. (\cite{ordal88}).  
We denote the optical depths of silicate and metallic iron grains 
as \mbox{$\tau_{V}^{\rm sil}$}\ and \mbox{$\tau_{V}^{\rm Fe}$}, respectively, and the exponents of the 
radial density distributions of both grain species are set to be equal 
and denoted as \mbox{$p$}.  
A grid of models is calculated for the following ranges 
of the parameters: \mbox{$\tau_{V}^{\rm sil}$}\ = 15 ... 35 with 
$\Delta \mbox{$\tau_{V}^{\rm sil}$} = 5$, \mbox{$\tau_{V}^{\rm Fe}$}\ = 1 ... 5 with 
$\Delta \mbox{$\tau_{V}^{\rm Fe}$} = 1$, \mbox{$r_{\rm in}$}\ = 10 ... 30~\mbox{$R_{\star}$}\ with 
$\Delta \mbox{$r_{\rm in}$} = 5$~\mbox{$R_{\star}$}, $\Theta$ = 10\degr\ ... 90\degr\ with 
$\Delta \Theta = 20\degr$, and \mbox{$p$}\ = 1.4, 1.6, 1.8, 2.0.  
The outer radius of the disk is set to $500 \times \mbox{$r_{\rm in}$}$.  

Figure~\ref{bestmodel_silfe_same} shows the best-fit model 
in this geometry.  The optical depths of silicate and 
iron dust are found to be $20 \pm 5$ and $3 \pm 1$ in the optical, 
respectively, 
which correspond to 10~\mbox{$\mu$m}\ optical depths of 1.5 and 0.34, 
respectively.  The disk has a half-opening angle of $50\degr \pm
10\degr$ and an inner boundary radius of $20 \pm 5$~\mbox{$R_{\star}$}, 
where the temperatures of silicate and iron grains 
reach $\sim$620~K, with \mbox{$p$}\ = $1.6 \pm 0.1$.  
The inclination angle is derived to be $30\degr \pm 10\degr$.  
Figure~\ref{bestmodel_silfe_same}a shows that 
the observed SED is fairly reproduced by the model, although the model 
predicts the near-infrared flux to be slightly lower than the observations.  
This discrepancy may be due to the uncertainty of 
the effective temperature of the central star as well as the 
simplifications adopted in our model such as the simple disk 
geometry with two grain species completely coexisting.  
Figures~\ref{bestmodel_silfe_same}c and 
~\ref{bestmodel_silfe_same}e--g reveal that the wavelength 
dependence of the $N$-band visibilities predicted by this model 
is also in fair agreement with the MIDI observations.  
In particular, the visibilities predicted for the baseline 
vector parallel to the symmetry axis of the disk (thick and thin 
dotted lines) can roughly reproduce the observed visibilities.  
However, the predicted visibilities are again somewhat 
too flat between 8 and 10~\mbox{$\mu$m}\ compared to those observed, 
as in the cases of the two models presented above.  
The intensity distributions of the silicate and metallic iron 
components show different wavelength dependences, and the flux 
contributions of these two components also vary in the $N$ band.  
As in the cases of the silicate+amorphous carbon and 
small+large silicate grain models, 
the combination of these factors results in a wavelength 
dependence of the $N$-band visibility totally different 
from that predicted by the models with silicate dust alone.  

The masses of silicate and metallic iron dust are derived to be 
$4.2 \times 10^{-3}$~\mbox{$M_{\sun}$}\ and $3.9 \times 10^{-4}$~\mbox{$M_{\sun}$}, 
with bulk densities of 3.7~g~cm$^{-3}$ and 7.9~g~cm$^{-3}$ 
assumed for silicate and metallic iron, respectively.  
The mass fraction of metallic iron is about 9\%, 
which is of the same order as the value 4\% derived by 
Kemper et al. (\cite{kemper02}) for OH127.8+0.0.  
As they mention, such mass fractions of metallic iron 
are consistent with the standard interstellar abundances 
of Fe and Si, which is not the case for models using spherical 
metallic iron grains.  In fact, we also calculated models 
consisting of silicate and spherical metallic iron grains 
with $a = 0.1$~\mbox{$\mu$m}\ and found a parameter set that can 
fairly reproduce the observed SED and $N$-band visibilities.  
However, the mass of metallic iron grains predicted by such a 
model is comparable to that of silicate dust (a mass fraction 
of metallic iron dust as high as $\sim$30\%) and, therefore, 
cannot easily be explained unless a remarkable overabundance 
of Fe is assumed as discussed in Kemper et al. (\cite{kemper02}) 
and Harwit et al. (\cite{harwit01}).

\section{Discussion}
\label{sect_discuss}

While we assumed a dust disk surrounding the 
carbon-rich AGB star (and also its putative low-luminosity companion) 
in our models, it is still controversial how oxygen-rich 
material is stored around silicate carbon stars.  
Based on the ISO observation of the silicate carbon star V778~Cyg,  
Yamamura et al. (\cite{yamamura00}) point out that such a circumbinary 
disk is exposed to direct radiation pressure from the central star 
with a luminosity of $\sim \!\! 10^4$~\mbox{$L_{\sun}$}\ and that the disk
cannot exist for a long time.  
They suggest that the oxygen-rich material is 
stored in the circumstellar disk around a companion, instead of 
a circumbinary disk, and that the silicate emission originates in the 
outflow from the circum-companion disk (see Fig.~5 
in Yamamura et al. \cite{yamamura00}).  
It should be noted here, however, that the infrared excess of V778~Cyg 
is much smaller than that of IRAS08002.  The silicate emission 
originating in the outflow from the circum-companion disk would 
not be sufficient to account for the infrared excess observed toward 
IRAS08002, because the solid angle subtended by such an outflow would not 
be very large, and consequently, the amount of dust in the outflow 
would be small.  The infrared excess expected from this scenario may 
be sufficient to explain the observed infrared excess of V778~Cyg, 
but possibly not that of IRAS08002.  

Also, the optical depths of the disk in our
models are significantly larger than that observed in V778~Cyg.  
Yamamura et al. (\cite{yamamura00}) stress that the silicate emission 
is optically thin, and the spherical shell modeling of Kwok \& Chan 
(\cite{kwok93}) also suggests that the optical depth of silicate grains in 
V778~Cyg is $\sim$0.1 at 10~\mbox{$\mu$m}, which is significantly 
smaller than the 1.5--1.9 that we derived for IRAS08002.  
In an optically thin case as in V778~Cyg, 
radiation pressure is likely to prevent a 
circumbinary disk from existing for an extended period of time.  
However, Yamamura et al. (\cite{yamamura00}) also point out that this
may not be the case for objects with optically thick disks such as 
the Red Rectangle and possibly the 
silicate carbon star IRAS09425-6040 with high crystallinity 
and that such optically thick disks can form around close binaries.  
In those optically thick cases, radiation pressure from the central 
primary star acts only on the inner surface of the disk and the 
rest of the disk can be shielded from direct radiation pressure.  
The stability of such optically thick disks may also 
provide an environment favorable for grains to grow in large 
sizes.  
Therefore, the results of our modeling for IRAS08002 are 
qualitatively consistent with the stability of the circumbinary disk.  
Furthermore, the total dust masses derived for our models 
are of the order of $10^{-3}$~\mbox{$M_{\sun}$}\ and 
remarkably larger than those derived for V778~Cyg (2--10$\times 
10^{-6}$~\mbox{$M_{\sun}$}\ by Yamamura et al. \cite{yamamura00}, 
$2 \times 10^{-7}$~\mbox{$M_{\sun}$}\ by Szczerba et al. \cite{szczerba05}), 
but close to the dust mass around IRAS09425-6040 ($\sim \!\! 2\times 
10^{-3}$~\mbox{$M_{\sun}$}) derived by Molster et al. (\cite{molster01}).  
The density distributions in our models ($p \simeq 1.6$--1.8) are 
also noticeably flatter than those expected for constant mass 
outflows.  Molster et al. (\cite{molster01}) derived such a 
flat density gradient to explain the ISO spectrum of IRAS09425-6040.  
These results imply that IRAS08002 may belong to the class of 
objects (presumably) possessing a dense dust disk like IRAS09425-6040 
and the Red Rectangle.  

We also mention that the inner boundary radii of our 
models are $\sim$15--20~\mbox{$R_{\star}$}, which means that 
the diameter of the inner cavity is $\sim$30--40~\mbox{$R_{\star}$}.  
If IRAS08002 is a binary consisting of a carbon-rich primary and a 
low-luminosity companion, the separation cannot be larger than the 
diameter of the inner cavity and is estimated to smaller than 
$\sim$30~\mbox{$R_{\star}$}\ ($\sim$60~AU).  This is in contrast with the 
separation between the binary components in the V778~Cyg system 
($\ga 75$~AU) estimated by Szczerba et al. (\cite{szczerba05}).  
Therefore, while such constraints are still rather weak, 
these results are at least qualitatively consistent 
with the scenario where a circumbinary disk may form in a close 
binary system and a circum-companion disk in a wide binary. 

We note that the silicate carbon star 
IRAS04496-6958 discovered in the Large Magellanic Cloud by 
Trams et al. (\cite{trams99}) exhibits a broad 10~\mbox{$\mu$m}\ spectrum 
similar to that of IRAS08002.  
Trams et al. (\cite{trams99}) suspect that 
the spectrum of IRAS04496-6958 may be explained by silicate and a 
second grain species such as large silicate grains, silicon carbide 
(SiC), crystalline olivines, and corundum.  As we mentioned in 
Sect.~\ref{sect_multgrain}, we did not consider crystalline olivines 
as the second grain species, because it is not (yet) identified 
in IRAS08002.  We also calculated disk models with SiC or corundum 
as the second grain species, using the same geometry where silicate 
and the second grain species coexist everywhere in the disk.  
However, as far as such a geometry is used, we could not find a model 
which reasonably reproduces the observed SED and $N$-band 
visibilities, because these models tend to show too strong features 
due to corundum (at $\sim$11--15~\mbox{$\mu$m}) or SiC 
(at $\sim$11.3~\mbox{$\mu$m}). 
Therefore, corundum or SiC may be present in the 
circumstellar environment of IRAS08002 as minor dust components, 
but probably not as major components responsible for the unexpected 
wavelength dependence of the $N$-band visibilities.  

There are still 
discrepancies between the observations and the models.  
In particular, 
all three models presented above show the same problem: 
the predicted $N$-band visibilities are too flat between 8 and 
10~\mbox{$\mu$m}, compared to the MIDI observations.  
This discrepancy may be attributed to the assumptions used in 
our models, such as the rather simple disk geometry and the simplified 
representation of the dust properties.  
Since the disk geometries considered in the present work are 
obviously quite simplified, more complicated disk geometries 
may at least partially explain the discrepancy between the 
observations and model predictions.  
An alternative explanation for this systematic discrepancy 
would be to assume that different grain species have different spatial 
distributions, instead of coexisting everywhere in the disk.  
It is plausible that distinct regions of the disk may be predominantly 
populated with different grain species due to the grain formation 
and growth processes.  
However, more observational constraints obtained by complementary 
observations would be 
needed to explore such complicated geometries and spatial 
variations of dust chemistry.

\section{Concluding remarks}

Our MIDI observations of IRAS08002 have resolved the dusty environment 
of a silicate carbon star for the first time.  
The observed visibilities show a monotonic 
increase from 8 to $\sim$9.5~\mbox{$\mu$m}, while they remain 
approximately constant longward of 10~\mbox{$\mu$m}.  
Although axisymmetric disk models consisting of silicate dust alone can 
reproduce the observed SED reasonably well, they 
fail to explain the observed $N$-band visibilities. 
This demonstrates the power of spectro-interferometry 
for probing the circumstellar environment of objects with a complex 
geometry and intricate dust chemistry.  

We then considered models with silicate and a second grain 
species, for which we adopted amorphous carbon, large silicate grains, 
and metallic iron grains.  
We have shown that such disk models can fairly reproduce the 
observed SED and $N$-band visibilities.  The dust disk around 
IRAS08002 is optically thick in all these 
models (\mbox{$\tau_{V}^{\rm sil}$}\ = 20--25) with inner radii of 15--20~\mbox{$R_{\star}$}\ 
and half-opening angles of 30\degr --50\degr.  
The derived density distributions ($ \rho \propto r^{-1.6}, r^{-1.8}$) 
are flatter than those expected for constant outflows.  
These results are qualitatively consistent with the scenario 
in which a dense circumbinary dust disk forms in some silicate carbon 
stars, possibly with a close companion.  
However, the wavelength 
dependence of the $N$-band visibilities predicted by these models  
is still somewhat too flat between 8 and 10~\mbox{$\mu$m}, as compared 
to the MIDI observations.  
It should also be noted that given the simplifications used in the models 
as well as the lack of complementary observational constraints, 
the two-grain-species models should be regarded 
as one of the possible scenarios to explain the unexpected 
$N$-band visibilities of IRAS08002 revealed with MIDI.  

Our MIDI observations of IRAS08002 and radiative transfer modeling 
have revealed that the silicate carbon star puzzle is even more 
puzzling than previously thought.  
Mid-infrared interferometric observations with shorter 
baselines and a wider coverage of position angles would be 
indispensable for probing the geometry of the dust disk.  
In order to investigate detailed dust chemistry, 
high-resolution mid-infrared spectroscopy would be crucial.  
Far-infrared and sub-mm observations would also be useful for detecting 
the presence or absence of large grains. 

\begin{acknowledgement}
The authors would like to thank T.~Kozasa and I.~Yamamura for 
valuable comments and discussions on silicate carbon stars as well as 
on dust formation in AGB stars.  
We are indebted to D.~Vinkovi\'c and 
A.~B.~Men'shchikov for fruitful discussions about radiative 
transfer calculations.  We also acknowledge D.~Engels for 
providing us with the results of the VLBA observations before
publication.  
\end{acknowledgement}

\end{document}